# Discretized Wiener–Khinchin theorem for Fourier–Laplace transformation: application to molecular simulations

Akira Koyama[1,2,a)], David A. Nicholson[2], Marat Andreev[2], and Gregory C. Rutledge[2], Koji Fukao[3], and Takashi Yamamoto[4]

[1]*Department of General Education, National Institute of Technology, Toyota College, 2-1 Eisei-cho, Toyota, Aichi 471-8525, Japan*
[2]*Department of Chemical Engineering, Massachusetts Institute of Technology, 77 Massachusetts Avenue, Cambridge, Massachusetts 02139, USA*
[3]*Department of Physical Sciences, Ritsumeikan University, Kusatsu, Shiga 525-8577, Japan*
[4]*Department of Physics and Informatics, Yamaguchi University, Yamaguchi 753-8512, Japan*

**ABSTRACT**

The Wiener–Khinchin theorem for the Fourier–Laplace transformation (WKT-FLT) provides a robust method to calculate numerically single-side Fourier transforms of arbitrary autocorrelation functions from molecular simulations. However, the existing WKT-FLT equation produces two artifacts in the output of the frequency-domain relaxation function. In addition, these artifacts are more apparent in the frequency-domain response function converted from the relaxation function. We find the sources of these artifacts that are associated with the discretization of the WKT-FLT equation. Taking these sources into account, we derive the new discretized WKT-FLT equations designated for both the frequency-domain relaxation and response functions with the artifacts removed. The use of the discretized WKT-FLT equations is illustrated by a flow chart of an on-the-fly algorithm. We also give application examples of the discretized WKT-FLT equations for computing dynamic structure factor and wave-vector-dependent dynamic susceptibility from molecular simulations.

[a)]Electronic mail: koyama@toyota-ct.ac.jp.

## I. INTRODUCTION

The Wiener–Khinchin theorem (WKT)[1,2] is a well-known general-purpose method to obtain Fourier transforms of arbitrary autocorrelation functions (ACFs) by computing the power spectral density. The theorem states that the power spectral density of a time series of a variable in a stationary state always exists even when the time series is neither absolutely integrable nor square integrable, and that the power spectral density coincides with the Fourier transform of the ACF. Originally, the WKT was developed and used to analyze data obtained from linear time-invariant systems constructed as electric circuits.[3] Today, the theorem plays important roles in various fields of signal processing and analysis.

In the fields of materials science and technology, the WKT is deeply connected with the fluctuation-dissipation theorem[4–7] and linear-response theory[8], and it contributes for the construction of stationary-state nonequilibrium physics and chemistry[9] as an essential tool. These theoretical approaches are widely used to investigate the dynamics observed in materials both experimentally and numerically. Recently, the WKT itself is also applied to obtain the Fourier transforms of ACFs and to investigate the molecular processes in numerical simulations. Based on developments in parallel computation tools such as multithreading, message-passing interfaces, general-purpose GPU calculations, and supported by high-performance solid-state memory devices, numerically computed spectroscopic data are now comparable to experimental data.[10–16] Some such data were computed by using the WKT instead of by the direct Fourier transformation for the ACFs. Here, we explain how to use the WKT in molecular simulations and present a powerful algorithm for implementing the WKT.

Given a physical quantity $f(t)$ as a function of time $t$, the ACF is

$$F(t) \equiv \frac{\langle f(t)f(0)\rangle}{\langle f(0)f(0)\rangle} \cong \frac{1}{A}\langle f(t+t_0)f(t_0)\rangle_{t_0} \;, \tag{1}$$

where $\langle \cdots \rangle$ means the statistical average and $A$ is a normalization factor. By assuming an ergodic system, we replace the statistical average $\langle \cdots \rangle$ with a long-time average $\langle \cdots \rangle_{t_0}$ over $t_0$.[9]

Both in deterministic and stochastic cases, *the WKT equation* is written as the following relation:

$$F(\omega) \cong I(\omega) \;, \tag{2}$$

where $\omega$ is an angular frequency, $F(\omega)$ is the Fourier transform of the ACF $F(t)$, and $I(\omega)$ is the power spectral density of the physical quantity of $f(t)$. The functions $F(\omega)$ and $I(\omega)$ are

defined as follows:

$$F(\omega) \equiv \int_{-\infty}^{\infty} F(t) e^{i\omega t} dt, \quad (3)$$

and

$$I(\omega) \equiv \frac{1}{A} \lim_{T \to \infty} \frac{1}{T} \left| \int_{-T/2}^{T/2} f(t) e^{i\omega t} dt \right|^2, \quad (4)$$

where $T$ is a time interval. The WKT equation is interpreted as follows: *the Fourier transform of the ACF $F(t)$ is proportional to the product of the Fourier transform of $f(t)$ and its complex conjugate.* In the literature on molecular simulations, the WKT equation is often written in the following form for simplicity[14]:

$$F(\omega) = \frac{1}{A} \lim_{T \to \infty} \frac{1}{T} \left| \int_{-T/2}^{T/2} f(t) e^{i\omega t} dt \right|^2. \quad (5)$$

To use the WKT in molecular simulations, the continuous WKT equation (5) must be discretized. By replacing the continuous $\omega$ with a discrete $\omega_n = 2\pi n/T$, we remove the notation for the operation $\lim_{T \to \infty}$ [hereafter, we refer to the $\lim_{T \to \infty}$ as "the limit on $T$"] from Eq. (5) and replace the integrals with discrete summations. As a result, we get *the discretized WKT* equation as follows:

$$F(\omega_n) = \frac{1}{AT} \left\langle \left| \sum_{m=-M/2+1}^{M/2} \Delta t f(t_m) e^{i\omega_n t_m} \right|^2 \right\rangle = \frac{\Delta t}{AM} \langle |f(\omega_n)|^2 \rangle, \quad (6)$$

where *m*, *n*, and *M* are integers, and $\Delta t$ is a time mesh. Because the limit on $T$ has been removed from Eq. (5), we add the brackets of the statistical average $\langle \cdots \rangle$ to Eq. (6). When we actually compute $F(\omega_n)$, we replace the statistical average $\langle \cdots \rangle$ with the average over the number of simulations performed under the same thermodynamic and computational conditions. The discretized Fourier transformation for $f(t_m)$ is written as

$$f(\omega_n) = \sum_{m=-M/2+1}^{M/2} f(t_m) e^{i\omega_n t_m} \ . \tag{7}$$

In Eqs. (6) and (7), the variables and constants are

$$t_m = m\Delta t, \quad -M/2 + 1 \leq m \leq M/2 \ , \quad M \text{ is even,}$$

$$\omega_n = n\Delta\omega, \quad \Delta\omega = \frac{2\pi}{M\Delta t}, \quad 1 \leq n \leq M/2 \ , \quad \text{and}$$

$$T = M\Delta t \ , \tag{8}$$

where the range of $n$ is half that of $m$ because of the Nyquist sampling theorem[3,17], and the maximum value of $\omega_n = \pi/\Delta t$ is the Nyquist frequency. The constant $\Delta\omega$ gives the lower resolution limit of $F(\omega_n)$. By using the discretized WKT equation, we can compute the Fourier transform $F(\omega_n)$ of the ACF, which is called the "frequency-domain correlation (or relaxation) function", from molecular simulations.

For example, density, mass flow, stress, or other physical quantities are functions of time in molecular simulations and can be written as $f(t_m)$ in Eqs. (6) and (7). If we have the complete time series $f(t_m)$ in computer storage, we can apply the discrete Fourier transformation of Eq. (7) to get the Fourier components $f(\omega_n)$. According to Eq. (6), multiplying $f(\omega_n)$ by its complex conjugate gives the frequency-domain correlation functions of density, mass flow, stress, and other physical quantities, without computing their ACFs and without implementing the direct Fourier transformation for the ACFs. The WKT provides a much easier way to obtain $F(\omega_n)$.

For high frequency phenomena, a large simulation time steps $M$ is not required. In this case, we use the WKT according to the procedures written in the preceding paragraph. Conversely, for low-frequency phenomena (for which $M$ is large), it becomes hard to output and store the value of $f(t_m)$ at every time step because the input-output time and total amount of the data become huge. In addition, it takes a long time to compute the Fourier transform $f(\omega_n)$ using Eq. (7) after the simulations. Moreover, it becomes harder when the time series contains the positions or velocities of the atoms in the system. How can we use the WKT for low-frequency phenomena? It is realized by thinning-out the angular frequency $\omega_n$ combined with an on-the-fly algorithm[18] developed by Matsui and co-workers.

We usually compare the simulation results with experimental results. If the experimental results are spectroscopic data of low-frequency phenomena such as complex dielectric constants or dynamic viscoelastic moduli, $F(\omega_n)$ need not be computed for all angular frequencies because the experimental data are often plotted vs the logarithm of frequency. We can thus thin out the unnecessary values of $F(\omega_n)$ and make space between the adjacent $\omega_n$'s. For example, the expression for $\omega_n$ in Eq. (8) can be changed to

$$\omega_n = 2^{n-1}\Delta\omega, \quad 1 \leq n \leq X, \quad M = 2^X . \tag{9}$$

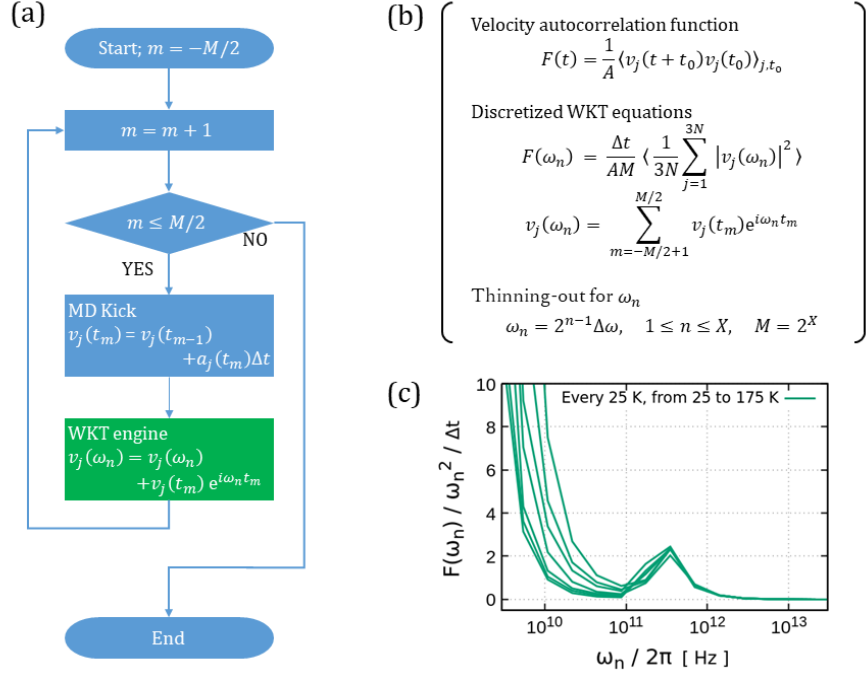

FIG. 1. On-the-fly algorithm for the WKT. (a) Flow chart of time loop of molecular dynamics simulation. (b) Equations for computing vibrational density of states $F(\omega_n)$. (c) Example of $F(\omega_n)$ vs log $\omega_n$. In panel (b), $v_j(t)$ is the velocity of the $j$th degree of freedom at time $t$. On the right-hand side of the equation for $F(\omega_n)$, a summation over $j$ is added where $N$ is the number of atoms. The angular frequency $\omega_n$ is thinned out by the equations on the last line in panel (b). When implementing the simulation of $2^{30}$ (>$10^9$) steps, it suffices to use $X = 30$ here. In panel (a), at the green-colored procedure named *WKT engine*, $v_j(t_m)e^{i\omega_n t_m}$ is computed at each time step and added sequentially to the array of $v_j(\omega_n)$. After multiple simulations, substituting $v_j(\omega_n)$ into the right-hand side of the equation for $F(\omega_n)$ and taking the average over the number of simulations gives the vibrational density of states $F(\omega_n)$. Panel (c) presents the results of a molecular dynamics simulation for the united atom polyethylene; $F(\omega_n)$ computed for the low-temperature glass is plotted vs the log $\omega_n$. The peak around $3 \times 10^{11}$ Hz is a "Boson peak", which appears in Raman or neutron scattering data.

By doing this, the memory that should be allocated for the array of $f(\omega_n)$ can be drastically reduced and a large array for $f(\omega_n)$ need not be set aside. For example, even for $F(\omega_n)$ from a simulation of $M = 2^{30}$ (> $10^9$) steps, preparing an array of $f(\omega_n)$ with 30 components suffices by setting $X = 30$ in Eq. (9) instead of the array with $\frac{1}{2} \times 2^{30}$ components according to Eq. (8). In addition, the summation over $m$ in Eq. (7) coincides with the time loop of the simulation. We can add a procedure to the simulation loop to implement the discrete Fourier transformation of Eq. (7) on the fly. In this procedure, the $m$th term of $f(t_m)e^{i\omega_n t_m}$ in the summation of Eq. (7) is computed during the $m$th simulation step and is added sequentially to the array of $f(\omega_n)$. Therefore, there is no need to output the entire time series of $f(t_m)$ during a simulation. After multiple simulations, substituting the computed $f(\omega_n)$ into Eq. (6), and taking the average over the number of the simulations, we get $F(\omega_n)$ without storing the entire time series of $f(t_m)$ and without any lack of $f(t_m)$. The post-simulation computation to get $F(\omega_n)$ is also reduced dramatically by this algorithm.

As an example, Fig. 1 shows a schematic illustration of how to compute the vibrational density of states $F(\omega_n)$ which is the Fourier transform of the velocity autocorrelation function of atoms. In Fig. 1, $f(t_m)$ and $f(\omega_n)$ from Eqs. (6) and (7) are replaced with the velocity $v_j(t_m)$ of the $j$th degree of freedom and its discrete Fourier transform $v_j(\omega_n)$, respectively. We call this approach the *on-the-fly algorithm for the WKT*.

However, the WKT has its limitations. The function $F(\omega_n)$ computed by using the WKT is always a real function because the ACF is a real, even function of time. Experimental data of spectroscopic measurements often have the imaginary part together with the real part; the former is related to energy loss, and the latter to energy stored. The complex frequency-domain correlation function is given by the Fourier–Laplace transformation (a single-side Fourier transformation) for the ACF, instead of by the Fourier transformation [Eq. (3)], which is written as

$$F^+(\omega) \equiv \int_0^\infty F(t)\, e^{i\omega t} dt \, , \qquad (10)$$

where the superscript $+$ ($-$) indicates that the interval of integration on the right-hand side is positive (negative). Of course, the imaginary part can be obtained analytically from the real part by using the Kramers–Kronig relation,[9] but this is difficult to do numerically. To obtain the imaginary parts from the simulations, we should modify and extend the usual WKT.

Matsui and co-workers proposed an extended WKT equation for the Fourier–Laplace transformation (WKT-FLT) to compute complex dynamic structure factor and complex dielectric constant from molecular simulations.[18,19] They applied the on-the-fly algorithm shown in Fig. 1 to the thinned-out WKT-FLT equations. Their strategy is robust and is attractive in many research fields, not only in materials physics but also in molecular chemistry and biology. However, their WKT-FLT equation produces two artifacts in $F^+(\omega_n)$ (Fig. 2), which become more apparent in the "frequency-domain response function" $\chi^+(\omega_n)$ (often called the "dynamic susceptibility"), which is converted from $F^+(\omega_n)$ by using[9]

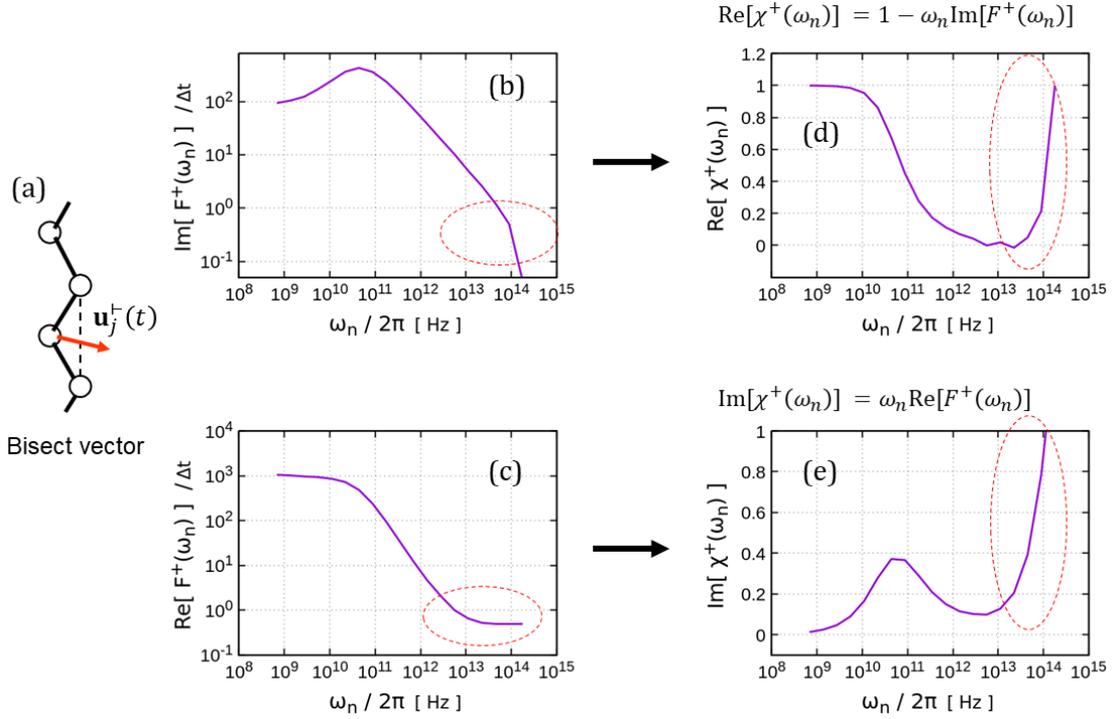

FIG. 2. Artefacts observed in $F^+(\omega_n)$ and $\chi^+(\omega_n)$ for the bisect vectors $\mathbf{u}_j^\vdash(t)$ [illustrated in panel (a)] of the united atom polyethylene model at 500 K, computed by using the existing WKT-FLT equation from molecular dynamics simulations. In the left column, panels (b) and (c) show the imaginary part and the real part of $F^+(\omega_n)$ plotted vs $\log \omega_n$, respectively. In the right column, (d) and (e) are the real part and the imaginary part of $\chi^+(\omega_n)$ converted from $F^+(\omega_n)$ by using $\chi^+(\omega) = 1 + i\omega F^+(\omega)$ [Eq. (11)]. The parts indicated by dashed red ellipses are the artifacts. In the left column, the upper curve should approach the asymptotic curve of $1/\omega_n$, and the lower curve should converge to zero, but they do not. The artifacts result in the strange increase in $\chi^+(\omega_n)$ for large $\omega_n$.

$$\chi^+(\omega) = 1 + i\omega F^+(\omega) \ . \tag{11}$$

In Figs. 2(d) and 2(e), both the real and imaginary parts of $\chi^+(\omega_n)$ are expected to converge to zero as $\omega_n$ increases. Thus, the imaginary part of $F^+(\omega_n)$ [Fig. 2(b)] should asymptotically approach $1/\omega_n$, and the real part of $F^+(\omega_n)$ [Fig. 2(c)] should converge to zero with increasing $\omega_n$. However, this does not happen: $\text{Im}[F^+(\omega_n)]$ deviates from $1/\omega_n$, and $\text{Re}[F^+(\omega_n)]/\Delta t$ converges to ½ for large $\omega_n$. The sources of this undesirable behavior remain unknown. We would like to clarify the sources and remove the artifacts. Thus, the goals of this study are

1. to clarify the sources of the artifacts, and
2. to remove the artifacts.

Depending on the research field, numerous names exist for the functions $F(t)$, $\chi(t)$, $F^+(\omega)$, and $\chi^+(\omega)$. To avoid confusion, we call $F(t)$ and $\chi(t)$ the time-domain relaxation function and response function, respectively, and we call their Fourier–Laplace transforms $F^+(\omega)$ and $\chi^+(\omega)$ the frequency-domain relaxation function and response function, respectively. The relaxation function is also referred to the correlation function, so we use the term "correlation" instead of relaxation if appropriate. In the time domain, we mainly deal with ACFs. Thus, in most cases, we call $F(t)$ the ACF.

Hereinafter, we use the notation $\text{Re}[\cdots]$ and $\text{Im}[\cdots]$ to refer to the real and imaginary parts of a complex quantity.

## II. EFFECTIVE AUTOCORRELATION FUNCTION

First, we derive an effective ACF for the WKT, which will be an important tool throughout this research.

Suppose that a quantity $f(t)$ of interest is a function of time $t$, then we define a long-time-averaged ACF $F(t)$ of $f(t)$ as

$$F(t) \equiv \frac{1}{A}\langle f(t+t_0)f(t_0)\rangle_{t_0} = \frac{1}{A}\lim_{T\to\infty}\frac{1}{T}\int_{-T/2}^{T/2} dt_0 \; f(t+t_0)f(t_0) \; . \tag{12}$$

The constant $A$ is a normalization factor defined as

$$A \equiv \langle f(t_0)f(t_0)\rangle_{t_0} = \lim_{T\to\infty}\frac{1}{T}\int_{-T/2}^{T/2} dt_0 \; f(t_0)f(t_0) \; . \tag{13}$$

Here, we assume that the time series of $f$ is computed from a molecular simulation limited to a finite time interval from $-T/2$ to $T/2$, and we use this $f$ for Eq. (12) after removing the limit on $T$. In this case, both $f(t+t_0)$ and $f(t_0)$ in Eq. (12) become restricted to the same time interval as the simulation. With this assumption, we restrict the domain of $f(t+t_0)$ to $-T/2 < t + t_0 < T/2$ as follows:

$$f(t+t_0) = \int_{-T/2}^{T/2} dt' \, f(t')\delta\bigl(t' - (t+t_0)\bigr) \; , \tag{14}$$

where $\delta(t)$ is the Dirac delta function. On substituting Eq. (14) into Eq. (12), we obtain the following form of *the effective ACF* $F_\text{e}(t)$:

$$F(t) \cong F_\text{e}(t) \equiv \frac{1}{A}\lim_{T\to\infty}\frac{1}{T}\int_{-T/2}^{T/2} dt_0 \int_{-T/2}^{T/2} dt' f(t')f(t_0) \; \delta\bigl(t' - (t+t_0)\bigr) \; . \tag{15}$$

Integrating Eq. (15) over $t'$ or $t_0$ with the cases $t' > t_0$ and $t' < t_0$, we see that $F_e(t)$ is an even function: $F_e(t) = F_e(-t)$.

The usual WKT equation is derived by using the effective ACF $F_e(t)$. Substituting $F_e(t)$ [Eq. (15)] into the right-hand side of Eq. (3) instead of $F(t)$ [Eq. (12)] gives

$$F(\omega) = \int_{-\infty}^{\infty} F_e(t)\, e^{i\omega t} dt$$

$$= \frac{1}{A} \lim_{T \to \infty} \frac{1}{T} \int_{-T/2}^{T/2} dt_0 \int_{-T/2}^{T/2} dt' f(t') f(t_0) \left\{ \int_{-\infty}^{\infty} dt\, e^{i\omega t} \delta(t' - (t + t_0)) \right\}$$

$$= \frac{1}{A} \lim_{T \to \infty} \frac{1}{T} \int_{-T/2}^{T/2} dt_0 \int_{-T/2}^{T/2} dt' f(t') f(t_0)\, e^{i\omega(t' - t_0)} . \tag{16}$$

We then obtain the WKT equation, which is the same as Eq. (5):

$$F(\omega) = \int_{-\infty}^{\infty} dt\, F_e(t)\, e^{i\omega t} = \frac{1}{A} \lim_{T \to \infty} \frac{1}{T} \left| \int_{-T/2}^{T/2} dt\, f(t)\, e^{i\omega t} \right|^2 . \tag{17}$$

Note that the effective ACF $F_e(t)$ [Eq. (15)] does not correspond exactly to $F(t)$ [Eq. (12)]. The integration area of Eq. (15) on the $t'$ vs $t_0$ plane differs from that of Eq. (12). See Appendixes A and B for details.

### III. WKT-FLT EQUATION FOR RELAXATION FUNCTION

This section derives a WKT-FLT equation with the effective ACF $F_e(t)$.

Introducing the Heaviside unit step function $\theta(t)$ into Eq. (10), $F^+(\omega)$ becomes

$$F^+(\omega) = \int_0^{\infty} dt\, F(t)\, e^{i\omega t} = \int_{-\infty}^{\infty} dt\, \theta(t) F(t)\, e^{i\omega t} , \tag{18}$$

where

$$\theta(t) \equiv \begin{cases} 1, & t > 0 \\ 1/2, & t = 0 \\ 0, & \text{otherwise} \end{cases} . \tag{19}$$

Note that the unit step function $\theta(t)$ is neither even nor odd. Thus, $\theta(t) F(t)$ in the integrand of Eq. (18) is also neither even nor odd. In Eq. (19), we set $\theta(0) = 1/2$ to satisfy the sum rules described later. Substituting the effective ACF $F_e(t)$ [Eq. (15)] into Eq. (18) instead of $F(t)$ [Eq. (12)], we obtain

$$F^+(\omega) = \int_{-\infty}^{\infty} dt\ \theta(t)F_e(t)\ e^{i\omega t}$$

$$= \frac{1}{A}\lim_{T\to\infty}\frac{1}{T}\int_{-T/2}^{T/2}dt_0\int_{-T/2}^{T/2}dt'\ f(t')f(t_0)\left\{\int_{-\infty}^{\infty}dt\ \theta(t)e^{i\omega t}\ \delta(t'-(t+t_0))\right\}$$

$$= \frac{1}{A}\lim_{T\to\infty}\frac{1}{T}\int_{-T/2}^{T/2}dt_0\int_{-T/2}^{T/2}dt'f(t')f(t_0)\ \theta(t'-t_0)\ e^{i\omega(t'-t_0)}$$

$$= \frac{1}{A}\lim_{T\to\infty}\frac{1}{T}\int_{-T/2}^{T/2}dt_0\int_{t_0}^{T/2}dt'f(t')f(t_0)\ e^{i\omega(t'-t_0)}\ . \tag{20}$$

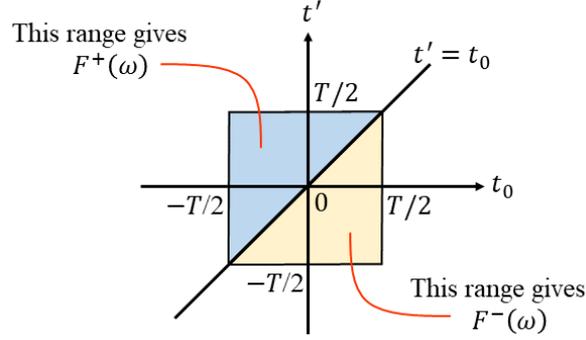

FIG. 3. Integration area of $F^+(\omega)$ and $F^-(\omega)$ over the $t'$ vs $t_0$ plane.

We then get *the WKT-FLT equation for the relaxation function*:

$$F^+(\omega) = \int_0^{\infty}dt\ F_e(t)\,e^{i\omega t} = \frac{1}{A}\lim_{T\to\infty}\frac{1}{T}\int_{-T/2}^{T/2}dt_0\ f(t_0)\,e^{-i\omega t_0}\int_{t_0}^{T/2}dt'f(t')\,e^{i\omega t'}\ . \tag{21}$$

Equation (21) is the same as the existing WKT-FLT equation.[18]

Next, we examine the WKT-FLT equation in detail. Introducing the unit step function $\theta(-t)$, $F^-(\omega)$ is written as

$$F^-(\omega) = \int_{-\infty}^{0}dt\ F(t)\,e^{i\omega t} = \int_{-\infty}^{\infty}dt\ \theta(-t)\ F(t)\,e^{i\omega t}. \tag{22}$$

In the same way as we derived Eq. (21), we get $F^-(\omega)$:

$$F^-(\omega) = \int_{-\infty}^{0} dt \ F_e(t) \, e^{i\omega t} = \frac{1}{A} \lim_{T \to \infty} \frac{1}{T} \int_{-T/2}^{T/2} dt_0 \ f(t_0) \, e^{-i\omega t_0} \int_{-T/2}^{t_0} dt' f(t') \, e^{i\omega t'}. \quad (23)$$

If we switch the order of the integrals in Eq. (23), switch the variables $t'$ and $t_0$, and take the complex conjugate, we see that

$$F^+(\omega) = \left(F^-(\omega)\right)^*, \quad (24)$$

where the notation $*$ means the complex conjugate. Combining Eqs. (21) and (23) gives

$$F(\omega) = F^+(\omega) + F^-(\omega), \quad (25)$$

where $F(\omega)$ is the Fourier transform of the ACF given in Eq. (5). Figure 3 shows the integration areas of $F^+(\omega)$ and $F^-(\omega)$ on the $t'$ vs $t_0$ plane. Even when integrating both sides of Eq. (25) over $\omega$, we still obtain

$$\int_{-\infty}^{\infty} d\omega \ F(\omega) = \int_{-\infty}^{\infty} d\omega \ F^+(\omega) + \int_{-\infty}^{\infty} d\omega \ F^-(\omega), \quad (26)$$

because the sum rules for $F(\omega)$, $F^+(\omega)$, and $F^-(\omega)$ are

$$\int_{-\infty}^{\infty} d\omega \ F(\omega) = \int_{-\infty}^{\infty} d\omega \int_{-\infty}^{\infty} dt \ F(t) e^{i\omega t} = 2\pi F(t=0), \quad (27)$$

and

$$\int_{-\infty}^{\infty} d\omega \ F^\pm(\omega) = \int_{-\infty}^{\infty} d\omega \int_{-\infty}^{\infty} dt \ \theta(\pm t) F(t) e^{i\omega t} = \pi F(t=0), \quad (28)$$

where the Dirac $\delta$ function is

$$\delta(t) = \int_{-\infty}^{\infty} \frac{d\omega}{2\pi} \, e^{i\omega t}. \quad (29)$$

Equation (26) holds because we set $\theta(0) = 1/2$. Otherwise, the right-hand side of Eq. (28) is not $\pi F(t=0)$, which invalidates Eq. (26). For Eq. (25) to safely hold, including the case of Eq. (26), we must define the unit step function as done in Eq. (19).

## IV. DISCRETIZED WKT-FLT EQUATION FOR RELAXATION FUNCTION AND A CORRECTION TERM FOR OVER-COUNTING

In this section, we discretize the WKT-FLT equation and show that the discretized version must include a correction term to eliminate an over-counting.

Replacing the continuous $\omega$ with the discrete $\omega_n = 2\pi n/T$ and removing the limit on $T$ from Eq. (21) gives

$$F^+(\omega_n) = \frac{1}{AT} \left\langle \int_{-T/2}^{T/2} dt_0 \ f(t_0) \ e^{-i\omega_n t_0} \int_{t_0}^{T/2} dt' f(t') \ e^{i\omega_n t'} \right\rangle , \qquad (30)$$

where the right-hand side of Eq. (30) is enclosed in brackets $\langle \cdots \rangle$. Because the statistical average is replaced with the long-time average when assuming ergodicity, we add the statistical-average brackets back to the equation when the limit on $T$ is removed. Again including the unit step function $\theta(t)$, we rewrite Eq. (30) in the form

$$F^+(\omega_n) = \frac{1}{AT} \int_{-T/2}^{T/2} dt_0 \int_{-T/2}^{T/2} dt' \theta(t' - t_0) \ \langle f(t_0)f(t') \rangle \ e^{i\omega_n(t'-t_0)} . \qquad (31)$$

Note that the integration in Eq. (31) over the line $t = t' - t_0 = 0$ contributes to the entire integration with *only half of its original value* because we have set $\theta(0) = 1/2$ for the sum rule for $F^\pm(\omega)$ to be satisfied as per Eq. (28). Consequently, Eq. (25) is confirmed to safely hold, including the case of Eq. (26).

Taking this into account, we now use the discretized form of the unit step function:

$$\theta_k = \begin{Bmatrix} 1 - \delta_{k,0}/2, & k \geq 0 \\ 0, & \text{otherwise} \end{Bmatrix} , \qquad (32)$$

where $\delta_{k,0}$ is the Kronecker delta and $k$ is an integer. Using $\theta_k$, Eq. (31) is discretized as

$$F^+(\omega_n) = \frac{1}{A}\frac{1}{M\Delta t}\sum_{l=-M/2+1}^{M/2}\sum_{m=-M/2+1}^{M/2}(\Delta t)^2\,\theta_{m-l}\,\langle f(t_l)f(t_m)\rangle\,e^{i\omega_n(t_m-t_l)}$$

$$= \frac{\Delta t}{AM}\sum_{l=-M/2+1}^{M/2}\sum_{m=l}^{M/2}\langle f(t_l)f(t_m)\rangle\,e^{i\omega_n(t_m-t_l)}$$

$$- \frac{\Delta t}{AM}\sum_{l=-M/2+1}^{M/2}\sum_{m=l}^{M/2}\langle f(t_l)f(t_m)\rangle\,e^{i\omega_n(t_m-t_l)}\left(\frac{1}{2}\delta_{m-l,0}\right). \tag{33}$$

For computational convenience, we switch the order of the double summation in the first term, which gives *the discretized WKT-FLT equation for the relaxation function*:

$$F^+(\omega_n) = \frac{\Delta t}{AM}\left\langle \sum_{m=-M/2+1}^{M/2} f(t_m)\,e^{i\omega_n t_m}\sum_{l=-M/2+1}^{m} f(t_l)\,e^{-i\omega_n t_l}\right\rangle - \frac{\Delta t}{2}, \tag{34}$$

where the variables and constants are the same as for Eq. (8). The second term is the correction term to eliminate the over-counting along the line $m = l$, which corresponds to the line $t = 0$ in the continuous case. The discretized function $F^-(\omega_n)$ is derived in the same way. The sum rules for the discretized WKT-FLT equations for $F^{\pm}(\omega_n)$ are easily confirmed by multiplying both sides of Eq. (34) by $\Delta\omega$ and summing from $n = -M/2 + 1$ to $M/2$ using the following representation of the Kronecker delta:

$$\delta_{m,l} = \frac{1}{M}\sum_{n=-M/2+1}^{M/2} e^{i\omega_n(t_m-t_l)}. \tag{35}$$

The relations involving $F(\omega_n)$, $F^+(\omega_n)$, and $F^-(\omega_n)$ hold in the same way as the relations for the continuous case [Eqs. (24)–(26)].

The discretized WKT-FLT equation can also be introduced into the simulation loop as done with the discretized WKT equation. The only difference with the case for the WKT is the green-colored procedure in Fig. 1. The discretized WKT-FLT equation (34) is separated as follows:

$$F^+(\omega_n) = \frac{\Delta t}{AM}\langle g(\omega_n)\rangle - \frac{\Delta t}{2}, \tag{36}$$

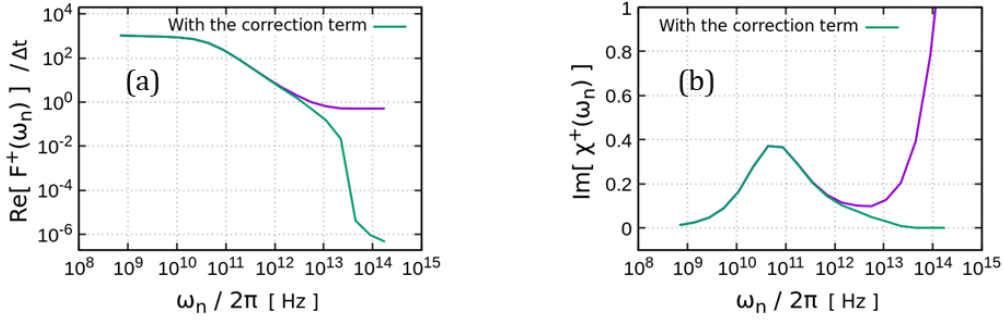

FIG. 4. (a) Real part of $F^+(\omega_n)$ and (b) imaginary part of $\chi^+(\omega_n)$ for the bisect vectors $\mathbf{u}_j^{\vdash}(t)$ [illustrated in Fig. 2(a)] of the united atom polyethylene model at 500 K, obtained by molecular dynamics simulation. The green and violet curves are computed by using the discretized WKT-FLT equation (34), with and without the correction term, respectively. The function $\chi^+(\omega_n)$ is converted from $F^+(\omega_n)$ by using Eq. (11). The artifacts observed in the violet curves are not present in the green curves.

$$g(\omega_n) = \sum_{m=-M/2+1}^{M/2} h(t_m, \omega_n) \ , \tag{37}$$

and

$$h(t_m, \omega_n) = f(t_m)\, e^{i\omega_n t_m} \sum_{l=-M/2+1}^{m} f(t_l)\, e^{-i\omega_n t_l} \ . \tag{38}$$

In Eq. (36), to compute $F^+(\omega_n)$, we replace the statistical average $\langle\cdots\rangle$ with the average over the number of the simulations.

Using Eq. (9), we thin out the unnecessary values of $F^+(\omega_n)$ and make space between the adjacent $\omega_n$'s. The value of $h(t_m, \omega_n)$ in Eq. (38) is computed every time step and is added to the array $g(\omega_n)$ of Eq. (37) during the simulation. After the end of the multiple simulations, substituting the computed $g(\omega_n)$ into Eq. (36) and taking the average over the number of simulations, we get $F^+(\omega_n)$ without storing the entire time series of $f(t_m)$ and without any lack of $f(t_m)$ during the simulations. We call this approach the *on-the-fly algorithm for the WKT-FLT*.

Figure 4 plots the real part of $F^+(\omega_n)$ and the imaginary part of $\chi^+(\omega_n)$ for the bisect vectors $\mathbf{u}_j^{\vdash}(t)$ [see Fig. 2(a)] as functions of $\log \omega_n$. The function $F^+(\omega_n)$ is computed by using the discretized WKT-FLT equation (34), and $\chi^+(\omega_n)$ is converted from $F^+(\omega_n)$ by using Eq. (11). Figure 4 shows both results computed with and without the correction term. Thanks to the correction term, the artifact is removed from $\text{Re}[F^+(\omega_n)]/\Delta t$ [Fig. 4(a)]: with the correction term, this quantity continues to decay with increasing $\omega_n$ (see green curve), whereas, without the correction term, it converges to 1/2 (see violet curve). As a result, the strange increase in Im[$\chi^+(\omega_n)$] for large $\omega_n$ is not present [Fig. 4(b)].

The source of the artifact in Re[$F^+(\omega_n)$] becomes clear; namely, it is due to the over-counting along the line $t = 0$ upon discretizing the WTK-FLT equation. Conversely, the source of the artifact in Im[$F^+(\omega_n)$] [Fig.2 (b)] remains unknown.

## V. SOURCE OF ARTIFACT IN Im[F⁺(ωₙ)]

As explained in the introduction, the real part of $\chi^+(\omega_n)$, converted from the imaginary part $F^+(\omega_n)$ by using Eq. (11) as Re[$\chi^+(\omega_n)$] = $1 - \omega_n$Im[$F^+(\omega_n)$], is expected to converge to zero with increasing $\omega_n$. Thus, Im[$F^+(\omega_n)$] should asymptotically approach $1/\omega_n$ with increasing $\omega_n$. Although Im[$F^+(\omega_n)$] in Fig. 2(b) appears to decrease proportionally to $1/\omega_n$ up to approximately $10^{13}$ Hz, it deviates from $1/\omega_n$ around $10^{14}$ Hz and drops to zero at $\omega_n = \pi/\Delta t$, where the $\pi/\Delta t$ is the maximum value of $\omega_n$ (called the Nyquist frequency).

We investigate this undesirable behavior of Im[$F^+(\omega_n)$] starting at Eq. (34). The equation for Im[$F^+(\omega_n)$] contains a factor of $\sin[\omega_n(t_m - t_l)]$ inside the double summation over $l$ and $m$. The value of $\sin[\omega_n(t_m - t_l)]$ is zero at the Nyquist frequency $\omega_n = \pi/\Delta t$, because $t_m - t_l = \Delta t(m - l)$ from Eq. (8). We expand it up to first order in a Taylor series in the vicinity of $\omega_n = \pi/\Delta t$. The resulting first-order term contains the factor $(\omega_n - \pi/\Delta t)$ outside the summation:

$$\mathrm{Im}[F^+(\omega_n)] = \frac{\Delta t}{AM} \sum_{m=-M/2+1}^{M/2} \sum_{l=-M/2+1}^{m} \langle f(t_m)f(t_l)\rangle \sin(\omega_n(t_m - t_l))$$

$$\cong \frac{\Delta t}{AM}\left(\omega_n - \frac{\pi}{\Delta t}\right) \sum_{m=-M/2+1}^{M/2} \sum_{l=-M/2+1}^{m} \langle f(t_m)f(t_l)\rangle (t_m - t_l)(-1)^{m-l}$$

$$= \frac{\Delta t}{A}\left(\omega_n - \frac{\pi}{\Delta t}\right) \sum_{k=-M/2+1}^{M/2} \theta_k \langle f(t_k)f(0)\rangle t_k(-1)^k$$

$$= \frac{\Delta t}{A}\left(\omega_n - \frac{\pi}{\Delta t}\right) \sum_{k=0}^{M/2} \langle f(t_k)f(0)\rangle t_k(-1)^k \quad , \tag{39}$$

where the double summation is changed to the single summation by using $\langle f(t_m)f(t_l)\rangle = \langle f(t_m - t_l)f(0)\rangle = \sum_{k=-M/2+1}^{M/2} \delta_{k,m-l} \langle f(t_k)f(0)\rangle$. Although it is difficult to know the precise value of the summation after the expansion, we consider, based on the simulation results shown in Fig. 2(b) and the model calculation in Appendix C, that it converges to a finite negative value. Thus, Im[$F^+(\omega_n)$] approaches $\omega_n = \pi/\Delta t$ proportionally to $-(\omega_n - \pi/\Delta t)$, and goes to zero at $\omega_n = \pi/\Delta t$. Accordingly, $\text{Re}[\chi^+(\omega_n)] = 1 - \omega_n \text{Im}[F^+(\omega_n)]$ increases with increasing $\omega_n$ around $\omega_n/2\pi = 10^{14}$ Hz and goes to unity at $\omega_n = \pi/\Delta t$, as in Fig. 2(d).

The Nyquist frequency $\omega_n = \pi/\Delta t$ does not appear in the finite-continuous Fourier transformation because the maximum value of $\omega_n$ (or $n$) can be set to $\pm\infty$. The Nyquist frequency only appears in the finite-discrete case. The finite-discrete $\text{Im}[F^+(\omega_n)]$ is always zero at $\omega_n = \pi/\Delta t$, and there is no way for $\text{Im}[F^+(\omega_n)]$ to approach $\omega_n = \pi/\Delta t$ except for $\text{Im}[F^+(\omega_n)] \propto -(\omega_n - \pi/\Delta t)$ near $\omega_n = \pi/\Delta t$. Therefore, the source of the undesirable behavior of $\text{Im}[F^+(\omega_n)]$ is the cutoff of the finite-discrete Fourier–Laplace transformation at the Nyquist frequency $\omega_n = \pi/\Delta t$. The source of the artifact in $\text{Im}[F^+(\omega_n)]$ clearly differs from that in $\text{Re}[F^+(\omega_n)]$ [see Sec. IV], although both artifacts are associated with the discretization of the WKT-FLT equation.

The Riemann–Lebesgue lemma states that the Fourier integral of an absolutely integrable function should converge to zero when $\omega$ or $\omega_n \to \infty$, whether the interval of the Fourier integral is finite or infinite.[17] It means that $F^+(\omega_n)$ and $\chi^+(\omega_n)$ analytically obtained by the finite-continuous Fourier-Laplace transformation converge to zero when $\omega_n \to 0$, because $F(t)$ and $\chi(t)$ are absolutely integrable. Hence, both the $F^+(\omega_n)$ and $\chi^+(\omega_n)$ computed from the finite-discrete molecular simulations should also converge to zero at large $\omega_n$. Actually, in Fig. 2(b), $\text{Im}[F^+(\omega_n)]$ partially reproduces the features of $\approx 1/\omega_n$. Moreover, when plotting $\text{Im}[F^+(\omega_n)]$ on the vertical axis instead of the logarithm of $\text{Im}[F^+(\omega_n)]$ [as in Fig. 2(b)], $\text{Im}[F^+(\omega_n)]$ appears to converge to zero. However, for the conversion of Eq. (11) not to produce the strange increase in $\text{Re}[\chi^+(\omega_n)]$, $\text{Im}[F^+(\omega_n)]$ must asymptotically approach $1/\omega_n$ but must not go to zero at $\omega_n = \pi/\Delta t$.

From the discussion here, it follows that the artifact observed in $\text{Im}[F^+(\omega_n)]$ is caused by the cutoff at the Nyquist frequency $\omega_n = \pi/\Delta t$. The way in which $\text{Im}[F^+(\omega_n)]$ approaches $\omega_n = \pi/\Delta t$ is different from the ideal case of $\approx 1/\omega_n$, which results in the strange increase of $\text{Re}[\chi^+(\omega_n)]$ near $\omega_n = \pi/\Delta t$ through the conversion $\text{Re}[\chi^+(\omega_n)] = 1 - \omega_n \text{Im}[F^+(\omega_n)]$ [Eq. (11)]. Although the reason is simple, it is difficult to eliminate the source from $\text{Im}[F^+(\omega_n)]$ and to remove the resulting artifact in $\text{Re}[\chi^+(\omega_n)]$.

Incidentally, $\chi^+(\omega)$ can be written as the Fourier–Laplace transform for a time-domain response function $\chi(t)$ instead of the conversion from $F^+(\omega)$ via Eq. (11). In this case, the discretized equation for $\text{Re}[\chi^+(\omega_n)]$ contains a correction term, same as that for $\text{Re}[F^+(\omega_n)]$ in Eq. (34). The correction term should eliminate the over-counting, allowing $\text{Re}[\chi^+(\omega_n)]$ to converge to zero for large $\omega_n$. In addition, the discretized equation for $\text{Im}[\chi^+(\omega_n)]$ contains the factor of $\sin[\omega_n(t_m - t_l)]$ inside the double summation, same as that for $\text{Im}[F^+(\omega_n)]$ in Eq. (39). Similarly to $\text{Im}[F^+(\omega_n)]$, $\text{Im}[\chi^+(\omega_n)]$ should appear to converge to zero at large $\omega_n$ as long as plotting the linearly separated $\text{Im}[\chi^+(\omega_n)]$ on the vertical axis, even though the asymptotic approach to the Nyquist frequency $\omega_n = \pi/\Delta t$ is not the ideal convergence. To remove the artifact, we must derive the discretized WKT-FLT equation designated for $\chi^+(\omega_n)$.

**VI. DISCRETIZED WKT-FLT EQUATION FOR RESPONSE FUNCTION**

This section derives the WKT-FLT equation for the response function $\chi^+(\omega)$ and then discretizes it.

The relation between the ACF $F(t)$ and the time-domain response function $\chi(t)$ is[9]

$$F(t) = \int_t^\infty \chi(s)\, ds = \int_{-\infty}^\infty \theta(s-t)\chi(s)\, ds, \tag{40}$$

where $\theta(s-t)$ is the Heaviside unit step function [Eq. (19)]. Differentiating Eq. (40) with respect to $t$ gives

$$\chi(t) = -\frac{dF(t)}{dt}, \tag{41}$$

where $d\theta(t)/dt = \delta(t)$ is used. Substituting the ACF of Eq. (12) into the above equation gives

$$\chi(t) = -\frac{1}{A}\langle \dot{f}(t+t_0)f(t_0)\rangle_{t_0} = -\frac{1}{A}\lim_{T\to\infty}\frac{1}{T}\int_{-T/2}^{T/2} dt_0\, \dot{f}(t+t_0)f(t_0), \tag{42}$$

where the dotted quantities ( $\dot{}$ ) are time derivatives. Following the strategy used to derive the effective ACF $F_e(t)$, we obtain an effective response function. We restrict the domain of $\dot{f}(t+t_0)$ to $-T/2 < t+t_0 < T/2$ as follows:

$$\dot{f}(t+t_0) = \int_{-T/2}^{T/2} dt'\, f(t')\frac{d}{dt}\delta\big(t'-(t+t_0)\big) = \int_{-T/2}^{T/2} dt'\, \dot{f}(t')\delta\big(t'-(t+t_0)\big). \tag{43}$$

Substituting Eq. (43) into Eq. (42), we get the *effective response function* $\chi_e(t)$:

$$\chi(t) \cong \chi_e(t) = -\frac{1}{A}\lim_{T\to\infty}\frac{1}{T}\int_{-T/2}^{T/2} dt_0 \int_{-T/2}^{T/2} dt'\, \dot{f}(t')f(t_0)\, \delta\big(t'-(t+t_0)\big). \tag{44}$$

The frequency-domain response function $\chi^+(\omega)$ is obtained by the Fourier–Laplace transformation for $\chi(t)$:

$$\chi^+(\omega) \equiv \int_0^\infty \chi(t)\, e^{i\omega t} dt = \int_{-\infty}^\infty \theta(t)\chi(t)\, e^{i\omega t} dt \ . \tag{45}$$

Substituting $\chi_e(t)$ [Eq. (44)] into Eq. (45) instead of $\chi(t)$ [Eq. (42)] gives *the WKT-FLT equation for the response function*:

$$\chi^+(\omega) = \int_0^\infty dt\, \chi_e(t)\, e^{i\omega t} = -\frac{1}{A}\lim_{T\to\infty}\frac{1}{T}\int_{-T/2}^{T/2} dt_0\, \dot{f}(t_0)\, e^{-i\omega t_0}\int_{t_0}^{T/2} dt' f(t')\, e^{i\omega t'} \ . \tag{46}$$

Next, we discretize Eq. (46) by replacing the continuous $\omega$ with the discrete $\omega_n = 2\pi n/T$ and removing the limit on $T$ from Eq. (46) by taking the statistical average $\langle \cdots \rangle$. We write it by using the unit step function $\theta(t)$:

$$\chi^+(\omega_n) = -\frac{1}{AT}\int_{-T/2}^{T/2} dt_0 \int_{-T/2}^{T/2} dt'\, \theta(t' - t_0)\, \langle \dot{f}(t_0) f(t')\rangle\, e^{i\omega_n(t'-t_0)} \ . \tag{47}$$

The integrals of Eq. (47) are discretized by using the discretized unit step function of Eq. (32):

$$\chi^+(\omega_n) = -\frac{1}{A}\frac{1}{M\Delta t}\sum_{l=-M/2+1}^{M/2}\sum_{m=-M/2+1}^{M/2}(\Delta t)^2\, \theta_{m-l}\, \langle \dot{f}(t_l) f(t_m)\rangle\, e^{i\omega_n(t_m - t_l)}$$

$$= -\frac{\Delta t}{AM}\sum_{l=-M/2+1}^{M/2}\sum_{m=l}^{M/2}\langle \dot{f}(t_l) f(t_m)\rangle\, e^{i\omega_n(t_m - t_l)}$$

$$+ \frac{\Delta t}{AM}\sum_{l=-M/2+1}^{M/2}\sum_{m=l}^{M/2}\langle \dot{f}(t_l) f(t_m)\rangle\, e^{i\omega_n(t_m - t_l)}\left(\frac{1}{2}\delta_{m-l,0}\right)$$

$$= -\frac{\Delta t}{AM}\sum_{l=-M/2+1}^{M/2}\sum_{m=l}^{M/2}\langle \dot{f}(t_l) f(t_m)\rangle\, e^{i\omega_n(t_m - t_l)} + \frac{\Delta t}{2AM}\sum_{l=-M/2+1}^{M/2}\langle \dot{f}(t_l) f(t_l)\rangle \ . \tag{48}$$

In Eq. (48), notice that the second term is zero because, at a given time $t_l$, a physical quantity $f(t_l)$ and its time derivative $\dot{f}(t_l)$ are uncorrelated [i.e., $\langle \dot{f}(t_l) f(t_l)\rangle = 0$].[9,20]

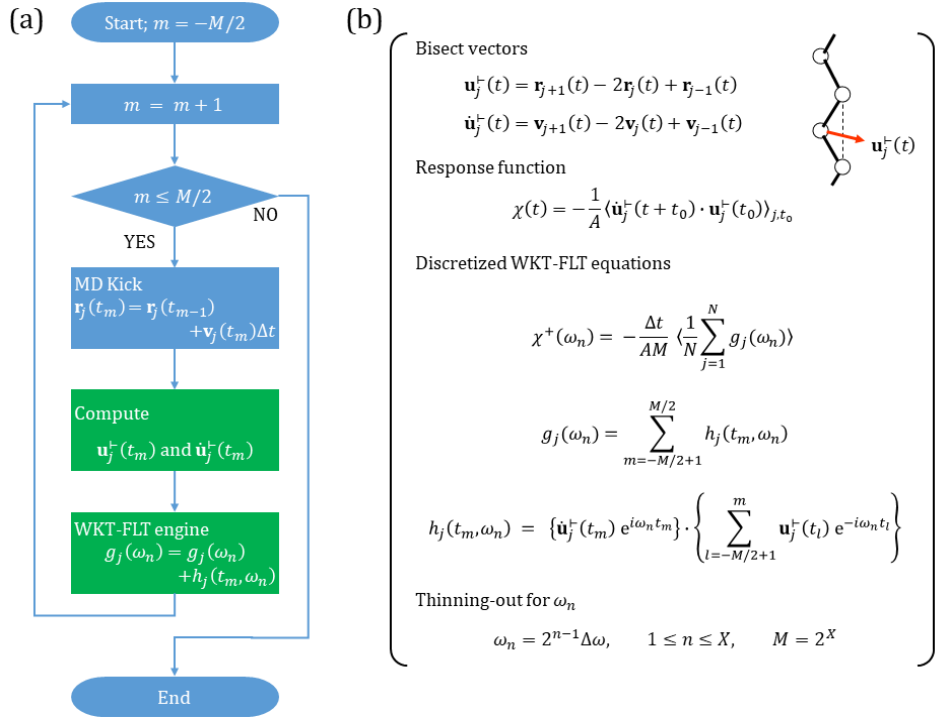

FIG. 5 On-the-fly algorithm for the WKT-FLT. (a) Flow chart of algorithm. (b) Equations for computing the frequency-domain response function $\chi^+(\omega_n)$ regarding the bisect vectors $\mathbf{u}_j^\vdash(t)$. Bisect vectors $\mathbf{u}_j^\vdash(t)$ and its time derivative $\dot{\mathbf{u}}_j^\vdash(t)$ are defined in the first and second equations in panel (b), where $\mathbf{r}_j(t)$ and $\mathbf{v}_j(t)$ are the position and velocity vectors of atom $j$, respectively. The time-domain response function $\chi(t)$, the discretized WKT-FLT equations for $\chi^+(\omega_n)$, and the equations for thinning out $\omega_n$ are also shown in panel (b). In panel (a), $\mathbf{u}_j^\vdash(t_m)$ and $\dot{\mathbf{u}}_j^\vdash(t_m)$ are computed by the first green-colored procedure. In the second green-colored procedure called *WKT-FLT engine*, $h_j(t_m, \omega_n)$ is computed at each time step and is added sequentially to the array of $g_j(\omega_n)$. After the end of the multiple simulations, substituting $g_j(\omega_n)$ into the first of the WKT-FLT equations and averaging over the number of simulations gives $\chi^+(\omega_n)$.

The final form of *the discretized WKT-FLT equation for the response function* is

$$\chi^+(\omega_n) = -\frac{\Delta t}{AM} \langle \sum_{m=-M/2+1}^{M/2} \dot{f}(t_m)\, e^{i\omega_n t_m} \sum_{l=-M/2+1}^{m} f(t_l)\, e^{-i\omega_n t_l} \rangle . \tag{49}$$

Contrary to our expectation in Sec. V, Eq. (49) does not have the correction term. Although the correction term to eliminate the over-counting along the line $m = l$ emerges in Eq. (48), it disappears due to the non-correlation between a physical quantity and its time derivative at a given time.

The discretized WKT-FLT equation for $\chi^+(\omega_n)$ can also be introduced into the simulation loop as done with $F^+(\omega_n)$. We separate the discretized WKT-FLT equation (49) as follows:

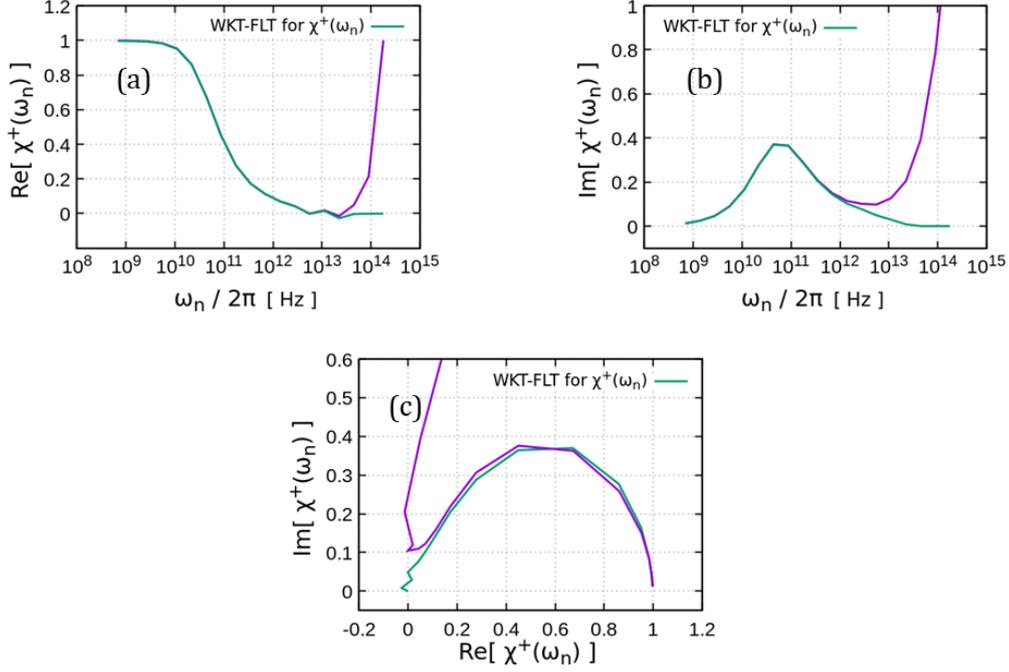

FIG. 6. $\chi^+(\omega_n)$ for bisect vectors of a united atom polyethylene model at 500 K, obtained by molecular dynamics simulation using the on-the-fly algorithm of Fig. 5. (a) Real part and (b) imaginary part of $\chi^+(\omega_n)$. (c) Cole–Cole plot of real and imaginary parts. The violet curves are obtained from the conversion from $F^+(\omega_n)$ using Eq. (11). The green curves are obtained from the designated WKT-FLT for the response function $\chi^+(\omega_n)$ [Eq. (49)]. In panels (a) and (b), the strange increases at large $\omega_n$ in the violet curves are not present in the green curves. Also, in panel (c), the error observed on the left-hand side of the half circle in the violet curve disappears in the green curve.

$$\chi^+(\omega_n) = -\frac{\Delta t}{AM} \langle g(\omega_n) \rangle ,  \qquad (50)$$

$$g(\omega_n) = \sum_{m=-M/2+1}^{M/2} h(t_m, \omega_n) , \qquad (51)$$

and

$$h(t_m, \omega_n) = \dot{f}(t_m) e^{i\omega_n t_m} \sum_{l=-M/2+1}^{m} f(t_l) e^{-i\omega_n t_l} . \qquad (52)$$

Using Eq. (9), we can thin out unnecessary values of $\chi^+(\omega_n)$. During the simulation, the function $h(t_m, \omega_n)$ of Eq. (52) is computed every time step $m$ and is added to the array $g(\omega_n)$ of Eq. (51). After the multiple simulations, we get $\chi^+(\omega_n)$ by substituting the computed $g(\omega_n)$ into Eq. (50) and taking the average over the number of the simulations. Figure 5 shows an outline of the on-the-fly algorithm for the WKT-FLT for a response function $\chi^+(\omega_n)$, which describes as an example the computation of the response function $\chi^+(\omega_n)$ for the bisect vectors $\mathbf{u}_j^\vdash(t)$ of a polymer chain.

Figures 6(a) and 6(b) plot the real and imaginary parts of $\chi^+(\omega_n)$ for the bisect vectors $\mathbf{u}_j^\vdash(t)$ computed by Eq. (49) (green curves) as functions of $\log \omega_n$. In addition, $\chi^+(\omega_n)$ converted from $F^+(\omega_n)$ by using Eq. (11) is also plotted (violet curves). The strange increases in both the real and imaginary parts of the violet curves at high $\omega_n$ are not present in the green curves. Figure 6(c) shows a Cole–Cole plot[21,22] for $\chi^+(\omega_n)$; the error on the left side of the half circle in the violet curve disappears in the green curve. Thus, the artifacts in both the real and imaginary parts of $\chi^+(\omega_n)$ are removed upon using the discretized WKT-FLT equation designated for $\chi^+(\omega_n)$.

## VII. APPLICATION EXAMPLES

General methods to measure directly the frequency-domain correlation functions of materials are the scattering experiments. For example, by using neutron[23–26] or x-ray[26,27] as a probe, we can obtain the frequency-domain density correlation function (called the dynamic structure factor) as a function of the wave vector and angular frequency from their inelastic scattering profiles. Or, using light, we can get the frequency-domain velocity correlation function (called the vibrational density of states) from the Raman scattering profile.[28–30] The vibrational density of states can also be obtained from the infrared absorption spectrum.[22,30] Conversely, by applying various external fields to sample materials, the spectroscopic data of the frequency-domain response functions are obtained. For example, we can measure the complex dielectric constants, complex magnetic susceptibilities or dynamic viscoelastic moduli[9,22,31–33] when applying electric, magnetic or mechanical fields to the materials, respectively.

All of these quantities are frequency-domain relaxation or response functions and can be computed, in principle, from molecular simulations by using the discretized WKT or WKT-FLT equations. Figure 1 already presents an example of the computation of the vibrational density of states and Fig. 5 shows the computation of the frequency-domain response function of the bisect vectors. In this section, starting from the van Hove time-space density correlation function for classical systems,[34] we derive the discretized WKT-FLT equations for the dynamic structure factor and the "wave-vector-dependent dynamic susceptibility".[9,18,35,36] For simplicity, we call these the frequency-domain density correlation function and response function, respectively.

The van Hove time-space density correlation function $G(\mathbf{r}, t)$ is defined as follows:

$$G(\mathbf{r}, t) \equiv \frac{1}{N}\langle \sum_{j=1}^{N}\sum_{k=1}^{N} \delta\left(\mathbf{r} + \mathbf{r}_j(0) - \mathbf{r}_k(t)\right) \rangle = \frac{1}{\rho}\langle \rho(\mathbf{r}, t)\rho^*(\mathbf{0}, 0) \rangle, \qquad (53)$$

where **r** is a position vector, $t$ is time, $N$ is the number of atoms, and $\mathbf{r}_j(t)$ is the position of atom $j$ at time $t$. The constant $\rho \equiv N/V$ is the average number density, where $V$ is the sample volume. The first form may be expressed in the second form[20] by using the expression for the density:

$$\rho(\mathbf{r}, t) = \sum_{j=1}^{N} \delta\left(\mathbf{r} - \mathbf{r}_j(t)\right) . \tag{54}$$

We replace the statistical average $\langle \cdots \rangle$ with the long-time average and space average $\langle \cdots \rangle_{t_0, \mathbf{r}_0}$ by assuming ergodicity:

$$\langle \cdots \rangle_{t_0} = \lim_{T \to \infty} \frac{1}{T} \int_{-T/2}^{T/2} dt_0 \ (\cdots) \ , \tag{55a}$$

$$\langle \cdots \rangle_{\mathbf{r}_0} = \frac{1}{V} \int_V d\mathbf{r}_0 \ (\cdots) = \lim_{L \to \infty} \frac{1}{L^3} \int_{-\mathbf{L}/2}^{\mathbf{L}/2} d\mathbf{r}_0 \ , \tag{55b}$$

where the integration over the volume $V$ in the first step of Eq. (55b) means that the integration of the position $\mathbf{r}_0$ encompasses all the sample material. Considering the application of the WKT-FLT to molecular simulations, we use the second form of Eq. (55b), and we use a cubic simulation box with periodic boundary: $\rho(\mathbf{r} \pm \mathbf{L}, t) = \rho(\mathbf{r}, t)$, where $\mathbf{L} = (L, L, L)$.

Using Eqs. (55a) and (55b), $G(\mathbf{r}, t)$ takes the form

$$G(\mathbf{r}, t) = \frac{1}{\rho} \langle \ \rho(\mathbf{r} + \mathbf{r}_0, t + t_0) \rho^*(\mathbf{r}_0, t_0) \ \rangle_{t_0, \mathbf{r}_0}$$

$$= \lim_{T, L \to \infty} \frac{1}{TN} \int_{-T/2}^{T/2} dt_0 \int_{-\mathbf{L}/2}^{\mathbf{L}/2} d\mathbf{r}_0 \ \rho(\mathbf{r} + \mathbf{r}_0, t + t_0) \rho^*(\mathbf{r}_0, t_0) \ , \tag{56}$$

where $N$ is included in the limit on $L$ because $N$ depends on $L$ for a constant $\rho$. In Eq. (56), the convolution over position $\mathbf{r}_0$ is written in the same way as the convolution over time $t_0$. However, they have different meanings when the limits on $T$ and $L$ are removed. For position, we impose the periodicity $\rho(\mathbf{r} \pm \mathbf{L}, t) = \rho(\mathbf{r}, t)$ in space with the periodic boundary, so that the value beyond the boundary can be replaced with the appropriate value from inside the simulation box. Conversely, for time, $\rho(\mathbf{r}, t)$ is limited to the fixed time range $|t| < T/2$, and no periodicity applies for $t$. We thus cannot apply the same replacement strategy for time as for position. Therefore, we must pay attention to the requirements of the WKT [see Appendixes A and B and Sec. VIII] in the case of time, whereas this is unnecessary in the case of space.

In Eq. (56), we thus restrict the time domain of $\rho(\mathbf{r} + \mathbf{r}_0, t + t_0)$ to $-T/2 < t + t_0 < T/2$, and impose the periodicity $\rho(\mathbf{r} \pm \mathbf{L}, t) = \rho(\mathbf{r}, t)$ as follows:

$$\rho(\mathbf{r} + \mathbf{r}_0 \pm \mathbf{L}, t + t_0) =$$
$$\rho(\mathbf{r} + \mathbf{r}_0, t + t_0) = \int_{-T/2}^{T/2} dt' \int_{-L/2}^{L/2} d\mathbf{r}' \rho(\mathbf{r}', t') \delta(\mathbf{r}' - (\mathbf{r} + \mathbf{r}_0)) \delta(t' - (t + t_0)). \tag{57}$$

Substituting Eq. (57) into Eq. (56), we obtain the effective time-space density correlation function $G_e(\mathbf{r}, t)$:

$$G(\mathbf{r}, t) \cong G_e(\mathbf{r}, t) = \lim_{T,L \to \infty} \frac{1}{TN} \int_{-T/2}^{T/2} dt_0 \int_{-T/2}^{T/2} dt' \int_{-L/2}^{L/2} d\mathbf{r}_0 \int_{-L/2}^{L/2} d\mathbf{r}' \rho(\mathbf{r}', t') \rho(\mathbf{r}_0, t_0)$$
$$\times \delta(\mathbf{r}' - (\mathbf{r} + \mathbf{r}_0)) \delta(t' - (t + t_0)) . \tag{58}$$

Next, substituting Eq. (58) into Eq. (41), we get the effective function $\chi_e(\mathbf{r}, t)$:

$$\chi(\mathbf{r}, t) \cong \chi_e(\mathbf{r}, t) = -\lim_{T,L \to \infty} \frac{1}{TN} \int_{-T/2}^{T/2} dt_0 \int_{-T/2}^{T/2} dt' \int_{-L/2}^{L/2} d\mathbf{r}_0 \int_{-L/2}^{L/2} d\mathbf{r}' \dot{\rho}(\mathbf{r}', t') \rho(\mathbf{r}_0, t_0)$$
$$\times \delta(\mathbf{r}' - (\mathbf{r} + \mathbf{r}_0)) \delta(t' - (t + t_0)) . \tag{59}$$

Note that $\chi(\mathbf{r}, t)$ and $\chi_e(\mathbf{r}, t)$ are response functions in time $t$ and are correlation functions in space $\mathbf{r}$.

By implementing the Fourier transformation in space and the Fourier–Laplace transformation in time for $G(\mathbf{r}, t)$ and $\chi(\mathbf{r}, t)$, we obtain the coherent (full-correlation) parts of the frequency-domain density correlation function $S^+(\mathbf{q}, \omega)$ and response function $\chi^+(\mathbf{q}, \omega)$, respectively, where $\mathbf{q}$ is the wave vector:

$$S^+(\mathbf{q}, \omega) = \int_0^\infty dt \int_V d\mathbf{r} \ G(\mathbf{r}, t) e^{-i(\mathbf{q} \cdot \mathbf{r} - \omega t)} = \int_{-\infty}^\infty dt \ \theta(t) \int_V d\mathbf{r} \ G(\mathbf{r}, t) e^{-i(\mathbf{q} \cdot \mathbf{r} - \omega t)} , \tag{60}$$

$$\chi^+(\mathbf{q}, \omega) = \int_0^\infty dt \int_V d\mathbf{r} \ \chi(\mathbf{r}, t) e^{-i(\mathbf{q} \cdot \mathbf{r} - \omega t)} = \int_{-\infty}^\infty dt \ \theta(t) \int_V d\mathbf{r} \ \chi(\mathbf{r}, t) e^{-i(\mathbf{q} \cdot \mathbf{r} - \omega t)} . \tag{61}$$

Substituting $G_e(\mathbf{r}, t)$ and $\chi_e(\mathbf{r}, t)$ into Eqs. (60) and (61) in the places of $G(\mathbf{r}, t)$ and $\chi(\mathbf{r}, t)$, respectively, we get the WKT-FLT equations for $S^+(\mathbf{q}, \omega)$ and $\chi^+(\mathbf{q}, \omega)$:

$$S^+(\mathbf{q}, \omega) = \lim_{T,L \to \infty} \frac{1}{TN} \sum_{j=1}^{N} \sum_{k=1}^{N} \int_{-T/2}^{T/2} dt_0 \int_{t_0}^{T/2} dt' \, \rho_j(\mathbf{q}, t') \rho_k^*(\mathbf{q}, t_0) e^{i\omega(t'-t_0)}$$

$$= \lim_{T,L \to \infty} \frac{1}{TN} \int_{-T/2}^{T/2} dt_0 \int_{t_0}^{T/2} dt' \, \rho(\mathbf{q}, t') \rho^*(\mathbf{q}, t_0) e^{i\omega(t'-t_0)} \, , \tag{62}$$

$$\chi^+(\mathbf{q}, \omega) = -\lim_{T,L \to \infty} \frac{1}{TN} \sum_{j=1}^{N} \sum_{k=1}^{N} \int_{-T/2}^{T/2} dt_0 \int_{t_0}^{T/2} dt' \{-i\mathbf{q} \cdot \dot{\mathbf{r}}_j(t)\} \rho_j(\mathbf{q}, t') \rho_k^*(\mathbf{q}, t_0) e^{i\omega(t'-t_0)}$$

$$= -\lim_{T,L \to \infty} \frac{1}{TN} \int_{-T/2}^{T/2} dt_0 \int_{t_0}^{T/2} dt' \, \dot{\rho}(\mathbf{q}, t') \rho^*(\mathbf{q}, t_0) e^{i\omega(t'-t_0)} \, , \tag{63}$$

where $\rho_j(\mathbf{q}, t)$ is the complex scattering amplitude of atom $j$:

$$\rho_j(\mathbf{q}, t) = \int_{-\mathbf{L}/2}^{\mathbf{L}/2} d\mathbf{r}' \delta(\mathbf{r}' - \mathbf{r}_j(t)) \, e^{-i\mathbf{q}\cdot\mathbf{r}'} = e^{-i\mathbf{q}\cdot\mathbf{r}_j(t)} \, . \tag{64}$$

The complex scattering amplitude of the whole system $\rho(\mathbf{q}, t)$ is the Fourier transform of the density $\rho(\mathbf{r}, t)$ [Eq. (54)] in space and is given by the superposition of $\rho_j(\mathbf{q}, t)$:

$$\rho(\mathbf{q}, t) = \int_{-\mathbf{L}/2}^{\mathbf{L}/2} d\mathbf{r}' \rho(\mathbf{r}', t) \, e^{-i\mathbf{q}\cdot\mathbf{r}'} = \sum_{j=1}^{N} e^{-i\mathbf{q}\cdot\mathbf{r}_j(t)} = \sum_{j=1}^{N} \rho_j(\mathbf{q}, t) \, . \tag{65}$$

By replacing $\omega$ with $\omega_n = 2\pi n/T$, removing the limits on $T$ and $L$, and adding the brackets $\langle \cdots \rangle$, Eqs. (62) and (63) are modified to the forms including the Heaviside unit step function $\theta(t' - t_0)$ of Eq. (19). In addition, they are discretized by replacing the integrals with the summations using the discretized unit step function $\theta_k$ of Eq. (32). After discretization, we get the following forms of the discretized WKT-FLT equations for the coherent (full-correlation) parts of the frequency-domain density correlation and response functions:

$$S^+(\mathbf{q_h}, \omega_n) = \frac{\Delta t}{MN} \langle \sum_{m=-M/2+1}^{M/2} \rho(\mathbf{q_h}, t_m) e^{i\omega_n t_m} \sum_{l=-M/2+1}^{m} \rho^*(\mathbf{q_h}, t_l) e^{-i\omega_n t_l} \rangle - \frac{\Delta t}{2} S(\mathbf{q_h}) , \quad (66)$$

$$\chi^+(\mathbf{q_h}, \omega_n) = -\frac{\Delta t}{MN} \langle \sum_{m=-M/2+1}^{M/2} \dot{\rho}(\mathbf{q_h}, t_m) e^{i\omega_n t_m} \sum_{l=-M/2+1}^{m} \rho^*(\mathbf{q_h}, t_l) e^{-i\omega_n t_l} \rangle , \quad (67)$$

where the second term in Eq. (66) is the correction term, and $S(\mathbf{q_h})$ is the discretized static structure factor:

$$S(\mathbf{q_h}) = \frac{1}{MN} \langle \sum_{m=-M/2+1}^{M/2} \rho(\mathbf{q_h}, t_m) \rho^*(\mathbf{q_h}, t_m) \rangle \quad (68)$$

In Eqs. (66)–(68), the wave vector $\mathbf{q}$ is discretized as

$$\mathbf{q_h} = (h_x \Delta q, h_y \Delta q, h_z \Delta q) , \quad \Delta q = \frac{2\pi}{L} , \quad h_x, h_y \text{ and } h_z = 1, 2, 3, \ldots . \quad (69)$$

The discretized forms of the incoherent (self-correlation) parts of the frequency-domain density correlation function $S_s^+(\mathbf{q_h}, \omega_n)$ and response function $\chi_s^+(\mathbf{q_h}, \omega_n)$ are obtained by introducing the Kronecker delta $\delta_{j,k}$ into the double summations of Eqs. (66)–(68) as follows:

$$S_s^+(\mathbf{q_h}, \omega_n) = \frac{\Delta t}{MN} \langle \sum_{j=1}^{N} \sum_{m=-M/2+1}^{M/2} \rho_j(\mathbf{q_h}, t_m) e^{i\omega_n t_m} \sum_{l=-M/2+1}^{m} \rho_j^*(\mathbf{q_h}, t_l) e^{-i\omega_n t_l} \rangle - \frac{\Delta t}{2} , \quad (70)$$

$$\chi_s^+(\mathbf{q_h}, \omega_n) = -\frac{\Delta t}{MN} \langle \sum_{j=1}^{N} \sum_{m=-M/2+1}^{M/2} \dot{\rho}_j(\mathbf{q_h}, t_m) e^{i\omega_n t_m} \sum_{l=-M/2+1}^{m} \rho_j^*(\mathbf{q_h}, t_l) e^{-i\omega_n t_l} \rangle . \quad (71)$$

Figure 7 plots the results of the real parts of $\chi_s^+(q_h, \omega_n)$ (upper row) and $\chi^+(q_h, \omega_n)$ (lower row) for the model polyethylene system as a function of the logarithm of $q_h$ and $\omega_n$. These functions are computed by combining the thinning-out of $\omega_n$ with the on-the-fly algorithm for the WKT-FLT. Because it takes a long time to compute $\chi^+(\mathbf{q_h}, \omega_n)$ and $\chi_s^+(\mathbf{q_h}, \omega_n)$ for all $\mathbf{q_h}$, we only compute the mean value of the following components of $\chi^+(\mathbf{q_h}, \omega_n)$ and $\chi_s^+(\mathbf{q_h}, \omega_n)$:

$$\chi^+(q_h, \omega_n) = \frac{1}{3}\{\chi^+((q_h, 0,0), \omega_n) + \chi^+((0, q_h, 0), \omega_n) + \chi^+((0,0, q_h), \omega_n)\},$$
$$q_h = h\Delta q, \qquad h = 1, 2, 3, \dots. \qquad (72)$$

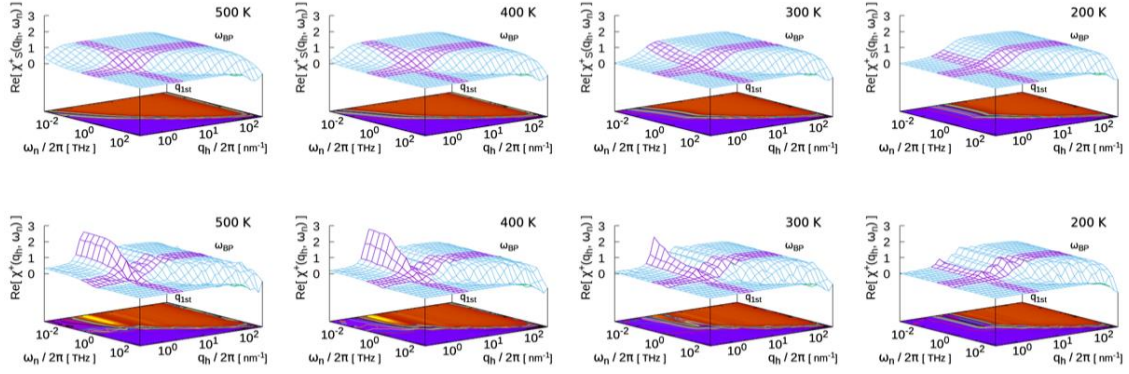

FIG. 7. Three-dimensional graphs of the real part of $\chi_s^+(q_h, \omega_n)$ and $\chi^+(q_h, \omega_n)$ computed for the model polyethylene system and plotted against the logarithm of $q_h$ and $\omega_n$. The upper and lower rows show $\chi_s^+(q_h, \omega_n)$ and $\chi^+(q_h, \omega_n)$, respectively. From left to right are shown the profiles for 500, 400, 300, and 200 K. The violet belts indicate the first peak position of $S(q_h)$ ($q_{first}$) and the Boson peak position ($\omega_{BP}$).

In the upper row of Fig. 7, the three-dimensional (3D) graphs change gradually with decreasing temperature. At 500 K, a single step appears connecting the low-$q_h$ and -$\omega_n$ region diagonally to the high-$q_h$ and -$\omega_n$ region. With decreasing temperature, the part of the step located around the low-$\omega_n$ range shifts to the high-$q_h$ range. At 200 K, the step in the low-$\omega_n$ range becomes parallel to the $q_h$ axis, whereas the step in the high-$q_h$ and -$\omega_n$ range remains diagonal, and the two steps intersect around $\omega_n/2\pi = 3.0 \times 10^{11}$ Hz $(\equiv \omega_{BP}/2\pi)$, which corresponds to the frequency of the Boson peak[23–25,27,37,38] shown in Fig. 1(C). Conversely, in the lower row, the 3D graphs are noisier than in the upper row, where the lower graphs look like that the upper graphs are multiplied by the static structure factor[39] $S(q_h)$ [Eq. (68)]. At high temperature, a strong streak appears on the line $q_h/2\pi = 2.0 \times 10^9$ m$^{-1}$ $(\equiv q_{1st}/2\pi)$ in the low-$\omega_n$ region, where the line corresponds to the first peak position of the $S(q_h)$. The streak shifts to the low-$\omega_n$ with decreasing temperature and goes out from the window at 200 K.

Figure 8 plots the results for the imaginary parts of $\chi_s^+(q_h, \omega_n)$ (upper row) and $\chi^+(q_h, \omega_n)$ (lower row). In the upper row, a single ridge extends diagonally from the low-$q_h$ and -$\omega_n$ region to the high-$q_h$ and -$\omega_n$ region. As the temperature decreases, the ridge located around the low-$\omega_n$ range shifts toward the high-$q_h$ range, as occurs for the real part in Fig. 7. The ridge in the low-$\omega_n$ region disappears at 200 K, where the remaining ridge terminates around $\omega_{BP}/2\pi$. In the lower row of Fig. 8, a strong peak appears on the same line $q_h/2\pi = q_{1st}/2\pi$ as for the strong streak in Fig. 7. As the temperature decreases, the peak shifts to the low-$\omega_n$ with changing its shape and is not present at 200 K.

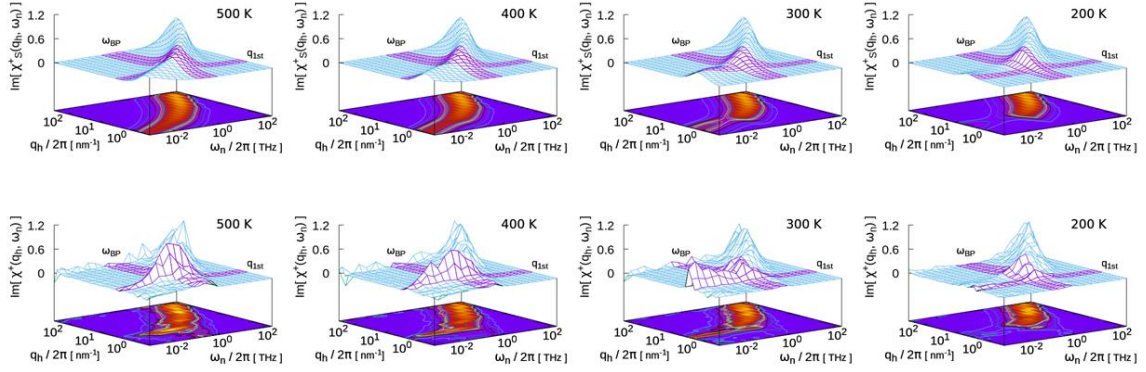

Fig. 8 Three-dimensional graphs of the imaginary part of $\chi_s^+(q_h, \omega_n)$ and $\chi^+(q_h, \omega_n)$ computed in the model polyethylene system are plotted against logarithm of $q_h$ and $\omega_n$. In the upper and lower lines, $\chi_s^+(q_h, \omega_n)$ and $\chi^+(q_h, \omega_n)$ are shown, respectively. From the left side to right side, the data of 500, 400, 300, and 200 K are aligned. The violet belts indicate the first peak positon of $S(q_h)$ ($q_{first}$), and the Boson peak's position ($\omega_{BP}$), respectively.

At 200 K in the upper rows, around the same $q_h/2\pi$ on the line $\omega_n/2\pi = \omega_{BP}/2\pi$, the two steps intersect and the ridge terminates; the $q_h/2\pi$ value might be link to the spatial scale of the Boson peak. Moreover, in the lower rows, the behavior of the strong streak and peak as a function of temperature should be connected to the glass transition around 200 K that occurs in the simulations based on polyethylene molecular models.[39–41]

## VIII. SUMMARY AND DISCUSSION

In Sec. II, we obtain the effective ACF $F_e(t)$ of Eq. (15) from the well-known time-averaged ACF $F(t)$ of Eq. (12) by restricting the time domain of the physical quantity $f(t + t_0)$ to $-T/2 < t + t_0 < T/2$. The effective ACF $F_e(t)$ is expressed by the double integral that includes the Dirac $\delta$ function. In Sec. III, We derive the WKT-FLT equation for the relaxation function $F^+(\omega)$ [Eq. (21)] by replacing the Fourier–Laplace (single-side) transformation for the ACF $F(t)$ with the Fourier (both-sides) transformation for the product of the effective ACF and the Heaviside unit step function: $\theta(t)F_e(t)$. Considering the sum rules of $F(\omega)$ and $F^\pm(\omega)$, we define the unit step function $\theta(t)$ in Eq. (19).

In Sec. IV, we discretize the continuous WKT-FLT equation, and notice that the integration along the line $t = t' - t_0 = 0$ contributes to the entire integral at only a half of its original value because we set $\theta(0) = 1/2$ for the sum rules of $F^\pm(\omega)$. To take this into account, we introduce the discretized unit step function $\theta_k$ [Eq. (32)], which is $1/2$ at $k = 0$. Upon using $\theta_k$, we find that the discretized WKT-FLT equation (34) for $F^+(\omega_n)$ must include a correction term to eliminate the over-counting along the line $m = l$, which corresponds to the line $t = 0$ in continuous case. As shown in Fig. 4, the artifact is removed from $\text{Re}[F^+(\omega_n)]$, and the strange increase in $\text{Im}[\chi^+(\omega_n)]$ disappears by the correction term. The source of the artifact in Re[$F^+(\omega_n)$] becomes clear; namely, it is due to the over-counting along the line $t = 0$ when we discretize the WTK-FLT equation.

Textbooks on signal processing and analysis[3] discuss the over-counting along the line $\omega_n = 0$ together with the over-counting along the Nyquist-frequency line $|\omega_n| = \pi/\Delta t$ in the case of inverse Fourier–Laplace transformations. The problem of over-counting is thus already recognized in the field of signal processing. However, Ref. 3 states that over-counting does not affect the computed results and can usually be ignored. In the present case, the artifact in Re[$F^+(\omega_n)$] is clearly connected to the over-counting along the line $t = 0$ and gives rise to the strange increase in Im[$\chi^+(\omega_n)$] in the high-frequency region as observed in Fig. 4. Therefore, in the present case, the over-counting produces a stronger effect than that in the case treated in Ref. 3. Conversely, we do not consider the over-counting along the edge of the integration interval of $|t| = T/2$ because the correlation between two values $f(t + t_0)$ and $f(t_0)$ for $|t| = T/2$ should be very weak if $T$ is sufficiently large.

Note that another artifact still remains in Im[$F^+(\omega_n)$] after removing the artifact due to the over-counting. We investigate the source in Sec. V, which clarifies that Im[$F^+(\omega_n)$] approaches the Nyquist frequency $\omega_n = \pi/\Delta t$ proportionally to $-(\omega_n - \pi/\Delta t)$, which is caused by the cutoff of the finite-discrete Fourier–Laplace transformation at the Nyquist frequency. The approach of Im[$F^+(\omega_n)$] to the Nyquist frequency differs from the ideal approach $\approx 1/\omega_n$, which results in the strange increase of Re[$\chi^+(\omega_n)$] near $\omega_n = \pi/\Delta t$ through the conversion $\chi^+(\omega) = 1 + i\omega F^+(\omega)$ [Eq. (11)]. The sources of the artifacts in Im[$F^+(\omega_n)$] and Re[$F^+(\omega_n)$] differ from each other, although both of them are associated with the discretization of the WKT-FLT equation of $F^+(\omega_n)$.

Pursuing the expectations based on the Riemann–Lebesgue lemma, we directly derive in Sec. VI the WKT-FLT equation designated for the response function $\chi^+(\omega_n)$ and discretize it. Contrary to our expectation, the discretized WKT-FLT equation (49) for the response function has no correction term. Although once the correction term for the over-counting along the line $m = l$ emerges in the discretized WKT-FLT equation, it disappears due to the lack of correlation between a physical quantity and its time derivative at a given time. Using the discretized WKT-FLT equation for the response function, we compute $\chi^+(\omega_n)$ for the bisect vectors of the polyethylene molecular model. The results [see Fig. 6] show that both artifacts previously observed in the real and imaginary parts of $\chi^+(\omega_n)$ are removed.

Incidentally, note that Im[$F^+(\omega_n)$] computed by Eq. (34) retains the artifact caused by the cutoff [see Fig. 2(b)] simply because its source has not been eliminated; the approach of Im[$F^+(\omega_n)$] to the Nyquist frequency still differs from the ideal case of $\approx 1/\omega_n$. However, when we plot linearly separated Im[$F^+(\omega_n)$] on the vertical axis [instead of logarithmically separated as in Fig. 2(b)], Im[$F^+(\omega_n)$] appears to converge to zero at large $\omega_n$. In Fig. 6(b), the same holds for Im[$\chi^+(\omega_n)$] (green curve) computed by Eq. (49). When we use the designated discretized WKT-FLT equations (34) and (49) to compute $F^+(\omega_n)$ and $\chi^+(\omega_n)$, the artifacts in their imaginary parts do not matter provided (i) we are not interested in how the functions approach the Nyquist frequency, and (ii) we plot the linearly separated Im[$F^+(\omega_n)$] and Im[$\chi^+(\omega_n)$] on the vertical axes.

Although we have not shown any results for the inverse conversion from $\chi^+(\omega_n)$ to $F^+(\omega_n)$ using $\chi^+(\omega) = 1 + i\omega F^+(\omega)$ [Eq. (11)], the function $F^+(\omega_n)$ converted from $\chi^+(\omega_n)$, which is computed by the designated equation for $\chi^+(\omega_n)$ [Eq. (49)], does not agree well at low or high $\omega_n$ with that computed by the designated equation for $F^+(\omega_n)$ [Eq. (34)]. We conclude that the discretized WKT-FLT equations (34) and (49) are not compatible with Eq. (11). To obtain the frequency-domain functions, the designated WKT-FLT equations are preferable, and it is better to avoid the conversion between $\chi^+(\omega_n)$ and $F^+(\omega_n)$ via Eq. (11).

The time-domain functions expressed by the double integral with the Dirac delta function play important roles in this research. In Secs. II, III, and VI, one may directly obtain the WKT or WKT-FLT equations by the Fourier or Fourier–Laplace transformation for the effective ACF $F_e(t)$ and the effective response function $\chi_e(t)$, respectively. Usually, the derivations of the WKT or WKT-FLT equations require several mathematical steps; however, thanks to the effective functions $F_e(t)$ and $\chi_e(t)$, such calculations are not required.

The WKT expresses the frequency-domain relaxation function $F(\omega_n)$ through the power spectral density $I(\omega_n)$; $F(\omega_n) \cong I(\omega_n)$. Conversely, the following relation expresses the time-domain relaxation function, ACF $F(t) = \frac{1}{A}\langle f(t)f(0)\rangle$, through the effective ACF $F_e(t)$:

$$\frac{1}{A}\langle f(t)f(0)\rangle \cong \frac{1}{A}\lim_{T\to\infty}\frac{1}{T}\int_{-T/2}^{T/2}dt_0\int_{-T/2}^{T/2}dt'\,f(t')f(t_0)\delta(t'-(t+t_0)). \tag{73}$$

The Fourier transform of Eq. (73) corresponds to the WKT equation (2), and the inverse Fourier transform of Eq. (2) corresponds to Eq. (73). In other words, Eq. (73) is the *time-domain WKT equation* as a counterpart of the frequency-domain WKT equation (2). Readers may already have noticed in Sec. II that the effective ACF $F_e(t)$ is the inverse Fourier transform of the power spectral density $I(\omega_n)$. Thus, Eq. (73) holds provided the WKT holds; the notation $\cong$ is thus valid when the requirements for the WKT are satisfied.

Statistical physics[9] textbooks state that, for the WKT equation (2) to hold, unnecessary terms should be omitted by taking the limit $T \to \infty$, where the convergences of the Fourier integrals of the ACF $F(t)$ and the function $tF(t)$ are required [see Appendix B]. Of course, the requirements for the WKT are correct in the infinite-continuous case. However, we must consider how to obtain reliable spectroscopic data from the finite-discrete molecular simulations. The discretization for the Fourier-transform equations might give rise to unexpected errors, as we have seen in this research. Therefore, more concrete and specific requirements are needed for the discretized WKT. Omitting the unnecessary terms[9] links to ignoring the highlighted areas with the broken lines on the plane of $t'$ vs. $t_0$ in Figs. A1(a) and A2(a). We now investigate the requirements for the WKT in various cases using the ACFs expressed by the double integrals on the $t'$ vs $t_0$ plane.

The time-domain WKT equation (73) can be applied to the Green–Kubo relation.[5,6,8] Integrating both sides of Eq. (73) over $t$ with an infinite integration interval gives

$$\frac{1}{A}\int_0^\infty dt\, \langle f(t)f(0)\rangle = \frac{1}{2A}\int_{-\infty}^\infty dt\, \langle f(t)f(0)\rangle$$

$$\cong \frac{1}{2AT}\int_{-T/2}^{T/2}dt_0 \int_{-T/2}^{T/2}dt'\, \langle f(t')f(t_0)\rangle \int_{-\infty}^\infty dt\, \delta(t'-(t+t_0))$$

$$= \frac{1}{2AT}\langle \left\{\int_{-T/2}^{T/2}dt\, f(t)\right\}^2 \rangle \,. \tag{74}$$

According to the Green–Kubo relation, the left-hand side of Eq. (74) gives the transport coefficient. The relation between the left-hand side of Eq. (74) and the final form on the right-hand side is discussed in the literature[8,42,43] and used to obtain arbitrary transport coefficients from molecular simulations. By using Eq. (8) instead of Eq. (9) to determine the variables in the last line of Fig. 1(b) and setting $\omega_n = 0$, we immediately obtain the *on-the-fly algorithm for the Green–Kubo relation*. For Fig. 1, the mobility (or diffusion coefficient) can be computed. Again, the notation $\cong$ in Eq. (74) is valid when the usual requirements for the WKT [Appendix B] are satisfied because we use the time-domain WKT in Eq. (74). However, the requirements for the discretized case are still unclear, which should also be known when using Eq. (74) to compute the transport coefficients from finite-discrete molecular simulations.

Although the time-domain WKT itself is not the subject of the present research, we confirm that both the frequency-domain WKT [Eq. (2)] and the Green–Kubo relation [Eq. (74)] can be derived by the Fourier integration over $t$ and by the integration over $t$, for the time-domain WKT [Eq. (73)], respectively. The frequency-domain WKT and the Green–Kubo relation are connected through the time-domain WKT. It would thus be of interest to further investigate the time-domain WKT.

In the introduction and in Sec. VI, we present the on-the-fly algorithm in the form of the flow charts in Figs. 1(a) and 5(a), originally developed by Matsui.[18] For example, in Sec. VI, we derive the discretized WKT-FLT equations for the dynamic structure factor $S^+(\mathbf{q_h}, \omega_n)$ and for the wave-vector-dependent dynamic susceptibility $\chi^+(\mathbf{q_h}, \omega_n)$. We also show the result of $\chi^+(q_h, \omega_n)$ computed by the WKT-FLT equation using the on-the-fly algorithm.

This algorithm can also be applied to obtain the frequency-domain relaxation and response functions for linearly separated $\omega_n$ instead of log-separated $\omega_n$. Thus, the algorithm works for computing high-frequency spectra comparable to experimental spectra, such as infrared absorption[14,30,44] and nuclear magnetic resonance[9,22,31] spectra. Conversely, the imaginary part is important, especially for low-frequency phenomena, where the data are plotted vs $\log \omega_n$ in most cases. Thus, the discretized WKT-FLT equations show their true strength when applied to the low-frequency phenomena using the on-the-fly algorithm with $\omega_n$ thinned out. One benefit of the on-the-fly algorithm is its simplicity, and the post-simulation calculation is reduced dramatically. We hope that the methods presented herein prove useful for investigating various molecular processes in material and biological applications.

**VIIII. CONCLUSIONS**

The two goals of this research are described in the introduction. The results of this research lead to the following conclusions:

1. We investigated the sources of the artifacts observed in the numerical results computed by the existing WKT-FLT equation for the frequency-domain relaxation function $F^+(\omega_n)$. The source of the artifact in $\text{Re}[F^+(\omega_n)]$ is the over-counting along the line $t = 0$ when the WKT-FLT equation is discretized. In contrast, the source of the artifact in $\text{Im}[F^+(\omega_n)]$ is the cutoff of the finite-discrete WKT-FLT equation at the Nyquist frequency. Through the conversion $\chi^+(\omega_n) = 1 + i\omega_n F^+(\omega_n)$, the artifacts yield similar strange increases in the real and imaginary parts of the frequency-domain response function $\chi^+(\omega_n)$ at high frequency. Although both of these sources are associated with the discretization of the WKT-FLT equation for $F^+(\omega_n)$, they differ from each other.

2. Taking the sources of the artifacts into account, we derived the new discretized WKT-FLT equation for $F^+(\omega_n)$ that included a correction term for the over-counting. Also, we derived the discretized WKT-FLT equation for $\chi^+(\omega_n)$. When using these corrected equations, the artifacts in $F^+(\omega_n)$ and $\chi^+(\omega_n)$ are removed.

3. The equation for the conversion $\chi^+(\omega) = 1 + i\omega F^+(\omega)$ is not compatible with the discretized WKT-FLT equations. Therefore, we recommend avoiding the conversion and instead computing $F^+(\omega_n)$ and $\chi^+(\omega_n)$ by using the equations designated for this purpose [Eqs. (34) and (49)].

4. We presented the on-the-fly algorithm for the WKT-FLT in the form of a flow chart. As examples, we derived the discretized WKT-FLT equations for the dynamic structure factor $S^+(\mathbf{q_h}, \omega_n)$ and the wave-vector-dependent dynamic susceptibility $\chi^+(\mathbf{q_h}, \omega_n)$. We also showed the computed results of $\chi^+(q_h, \omega_n)$.

More concrete and specific requirements for the WKT should become known when we apply it to molecular simulations. If successful, we should be able to use with confidence not only the WKT but also the WKT-FLT and the Green–Kubo relation for molecular simulations. Such research will be the subject of future presentations.

## ACKNOWLEDGMENTS

This research was supported by funding from National Institute of Technology, JAPAN, for the overseas training program. We would like to express our gratitude to KIOXIA Co. for providing NVMe solid-state drives for this research. Also, we would like to thank Professor Takashi Odagaki and Dr. Jun Matsui for their helpful comments.

## DATA AVAILABILITY

The data that support the findings of this study are available from the corresponding author upon reasonable request.

**APPENDIX A: DIFFERENCE BETWEEN EFFECTIVE ACF AND LONG-TIME AVERAGED ACF**

The effective ACF $F_e(t)$ of Eq. (15) is not the same as the well-known long-time-averaged ACF $F(t)$ of Eq. (12). Here, we show explicitly how they differ.

When the time series of $f$ is limited to a finite time interval from $-T/2$ to $T/2$ [as in Sec. II], the range of the time lag $t$ between $f(t+t_0)$ and $f(t_0)$ can be up to twice as large as the time interval. Thus, the domain of the time-averaged ACF is often set to be twice as large as the time interval. Here, as per convention, we set the time domain of $F_e(t)$ and $F(t)$ to $-T < t < T$.

First, we rewrite the effective ACF $F_e(t)$ [Eq. (15)] in the single-integral form and compare it with the long-time-averaged ACF $F(t)$ [Eq. (12)]. To do this, we define a window function $W[a < t < b]$ as follows:

$$W[a < t < b] \equiv \left\{ \begin{array}{ll} 1, & a < t < b \\ 0, & \text{otherwise} \end{array} \right\}. \tag{A1}$$

We insert the window function $W[-T/2 < t' < T/2]$ into the integral of Eq. (15) and remove the limit on $T$ from Eq. (15). However, we do not add the brackets $\langle \cdots \rangle$ for statistical averaging even after removing the limit on $T$, because the goal is to compare $F_e(t)$ with $F(t)$ on the planes $t$ vs $t_0$ and $t'$ vs $t_0$ depicted in Figs. A1 and A2, respectively. In this appendix, Eqs. (12) and (15) are treated as not having both the limit on $T$ and the statistical-averaging brackets $\langle \cdots \rangle$ (the necessity to do this is explained later). Then, we split the integral over $t'$ into two integrals of the intervals $-\infty < t' \le t_0$ and $t_0 \le t' < \infty$ by using the Heaviside unit step functions $\theta(t' - t_0)$ and $\theta(t_0 - t')$, and we integrate them over $t'$. The resulting effective ACF expressed by the single integral over $t_0$ is

$$F_e(t) = \frac{1}{AT} \int_{-T/2}^{T/2} dt_0 \int_{-\infty}^{\infty} dt' W[-T/2 < t' < T/2] \; f(t')f(t_0) \; \delta(t' - (t + t_0))$$

$$= \frac{1}{AT} \int_{-T/2}^{T/2} dt_0 \int_{-\infty}^{\infty} dt' \{\theta(t' - t_0) + \theta(t_0 - t')\}$$
$$\times W[-T/2 < t' < T/2] \; f(t')f(t_0) \delta(t' - (t + t_0))$$

$$= \frac{1}{AT} \{\theta(t) + \theta(-t)\} \int_{-T/2}^{T/2} dt_0 \; W[-T/2 < t + t_0 < T/2] \; f(t + t_0)f(t_0)$$

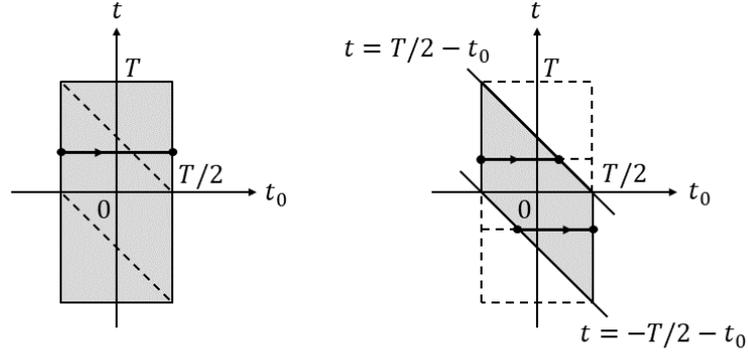

(a) Long-time averaged ACF $F(t)$     (b) Effective ACF $F_e(t)$

FIG. A1. Schematic diagrams of integration over $t_0$ in Eqs. (12) and (A2): (a) $F(t)$ and (b) $F_e(t)$. The integral over $t_0$ with a fixed $t$ means the integration along the line segment between two filled circles.

$$= \begin{cases} \dfrac{1}{AT}\int_{-T/2}^{T/2-t} dt_0 f(t+t_0)f(t_0), & t > 0 \\ 1, & t = 0 \\ \dfrac{1}{AT}\int_{-T/2-t}^{T/2} dt_0 f(t+t_0)f(t_0), & t < 0 \end{cases}. \quad (A2)$$

Figure A1 shows schematic diagrams of the integration area for $F(t)$ [Eq. (12)] and $F_e(t)$ [Eq. A2)] on the plane $t$ vs $t_0$. The integral over $t_0$ with a fixed $t$ means integrating along the line segment between two filled circles. Figures A1(a) and A1(b) show that the two integration areas in $F(t)$ highlighted by the dashed lines are missing in $F_e(t)$. Note that, in the last line of Eq. (A2), the integration intervals over $t_0$ depend on $t$, although the integrals are divided by the constant $T$. Thus, $F_e(t)$ is not a time-averaged autocorrelation function in the strict sense.

Next, we rewrite the long-time-averaged ACF $F(t)$ of Eq. (12) in double-integral form, similar to $F_e(t)$ of Eq. (15). To restrict the time range of $t$ to $-T < t < T$, we write $f(t+t_0)$ with the window function $W[-T < t < T]$ as

$$f(t+t_0) = W[-T < t < T]\int_{-\infty}^{\infty} dt' f(t')\delta(t' - (t+t_0)). \quad (A3)$$

Substituting Eq. (A3) into Eq. (12) gives

$$F(t) = \frac{1}{AT}\int_{-T/2}^{T/2} dt_0 \int_{-\infty}^{\infty} dt' W[-T < t' - t_0 < T] f(t')f(t_0) \delta(t' - (t+t_0))$$

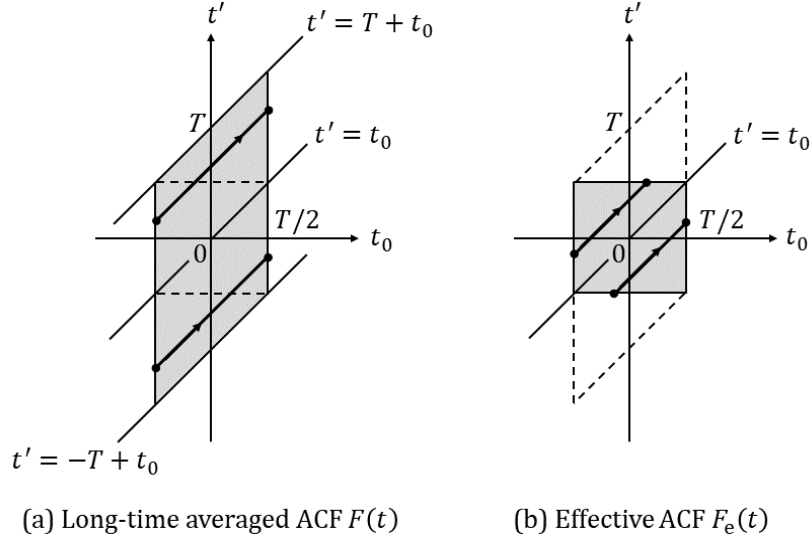

(a) Long-time averaged ACF $F(t)$  (b) Effective ACF $F_e(t)$

FIG. A2. Schematic diagrams of integration over $t'$ and $t_0$ in Eqs. (A4) and (15): (a) $F(t)$ and (b) $F_e(t)$. The double integral of Eqs. (A4) and (15) over $t'$ and $t_0$ means the integration along the line segment between two filled circles for a fixed $t$. The $t$ value is equal to the $t'$ value at the intercept of the line segment on the $t'$ axis.

$$= \frac{1}{AT} \int_{-T/2}^{T/2} dt_0 \int_{-T+t_0}^{T+t_0} dt' f(t') f(t_0) \, \delta(t' - (t + t_0)) \, . \tag{A4}$$

Figure A2 shows schematic diagrams of the integration areas of $F(t)$ [Eq. (A4)] and $F_e(t)$ [Eq. (15)] on the plane $t'$ vs $t_0$. The vertical axes are changed from $t$ in Fig. A1 to $t'$ in Fig. A2. In Fig. A2, the double integral of Eqs. (A4) and (15) over $t'$ and $t_0$ with the delta function $\delta(t' - (t + t_0))$ means that $f(t')f(t_0)$ is integrated along the line segment between two filled circles for a fixed $t$; the $t$ value is equal to the $t'$ value at the intercept of the line segment on the $t'$ axis. Figure A2 shows that two integration areas highlighted by the broken lines in $F(t)$ are missing in $F_e(t)$, which is the same as in Fig. A1.

If we add the statistical-averaging brackets $\langle \cdots \rangle$ to the last line of Eq. (A2), for example, the integration variable $t_0$ disappears:

$$F_e(t) = \frac{1}{AT} \int_{-T/2}^{T/2-t} dt_0 \langle f(t + t_0) f(t_0) \rangle = \frac{1}{AT} \left( \int_{-T/2}^{T/2-t} dt_0 \right) \langle f(t) f(0) \rangle$$

$$= \frac{1}{AT} (T - t) \langle f(t) f(0) \rangle, \quad \text{for } 0 < t \, , \tag{A5}$$

where $\langle f(t + t_0) f(t_0) \rangle = \langle f(t) f(0) \rangle$ is used. As a result, we can no longer compare the effective ACF $F_e(t)$ with the long-time-averaged ACF $F(t)$ on the planes $t$ vs $t_0$ and $t'$ vs $t_0$, as done in Figs A1 and A2. Thus, the brackets were not added to the ACFs even after removing the limit on $T$.

**APPENDIX B: REQUIREMENT FOR WKT**

In this appendix, we derive the requirement for the infinite-continuous WKT [Eq. (2)].

We define an ACF $F_1(t)$ over the highlighted areas outlined by dashed lines in Figs. A1 and A2. For $0 < t < T/2$, $F_1(t)$ is

$$F_1(t) \equiv \frac{1}{At} \int_{-T/2}^{T/2} dt_0 \int_{T/2}^{T+t_0} dt' f(t') f(t_0) \, \delta(t' - (t + t_0))$$

$$= \frac{1}{At} \int_{T/2-t}^{T/2} dt_0 \, f(t + t_0) f(t_0) \, , \tag{B1a}$$

and for $-T/2 < t < 0$,

$$F_1(t) \equiv \frac{1}{A(-t)} \int_{-T/2}^{T/2} dt_0 \int_{-T+t_0}^{-T/2} dt' f(t') f(t_0) \, \delta(t' - (t + t_0))$$

$$= \frac{1}{A(-t)} \int_{-T/2}^{-T/2-t} dt_0 \, f(t + t_0) f(t_0) \, . \tag{B1b}$$

The relationship between $F(t)$ of Eq. (A4), $F_e(t)$ of Eq. (A2), and $F_1(t)$ is

$$F_e(t) = \begin{cases} F(t) - \dfrac{t}{T} F_1(t) \, , & \text{for } t > 0 \\ F(t) + \dfrac{t}{T} F_1(t) \, , & \text{for } t < 0 \end{cases} = F(t) - \text{sgn}(t) \frac{t}{T} F_1(t) \, , \tag{B2}$$

where $\text{sgn}(t) \equiv \theta(t) - \theta(-t)$ is the sign function, and $\theta(t)$ is the unit step function defined in Eq. (19). When we add the brackets $\langle \cdots \rangle$ to Eq. (B2), $\langle F_1(t) \rangle$ on the right-hand side becomes equal to $\langle F(t) \rangle$:

$$\langle F_1(t) \rangle = \frac{1}{At} \int_{T/2-t}^{T/2} dt_0 \, \langle f(t+t_0) f(t_0) \rangle = \frac{1}{At} \left( \int_{T/2-t}^{T/2} dt_0 \right) \langle f(t) f(0) \rangle$$

$$= \frac{1}{A} \langle f(t) f(0) \rangle = F(t) \text{ for } t > 0, \tag{B3}$$

where, according to Eq. (1), we have removed the brackets from $F(t)$ in the last line of Eq. (B3). Equation (B2) then becomes

$$\langle F_\text{e}(t)\rangle = F(t) - \text{sgn}(t)\frac{t}{T}F(t) . \tag{B4}$$

Implementing the Fourier integrals on both sides of Eq. (B4) with the integration intervals $0 < t < T$ and $-T < t < 0$ gives

$$\int_0^T dt\ \langle F_\text{e}(t)\rangle e^{i\omega t} = \int_0^T dt\ F(t)e^{i\omega t} - \frac{1}{T}\int_0^T dt\ tF(t)e^{i\omega t} , \tag{B5}$$

and

$$\int_{-T}^0 dt\ \langle F_\text{e}(t)\rangle e^{i\omega t} = \int_{-T}^0 dt\ F(t)e^{i\omega t} + \frac{1}{T}\int_{-T}^0 dt\ tF(t)e^{i\omega t} . \tag{B6}$$

If we take the limit $T \to \infty$ in Eqs. (B5) and (B6), the second terms on the right-hand sides go to zero provided the Fourier integrals of $tF(t)$ converge to finite values for arbitrary $\omega$. In addition, as long as the first terms converge to finite values for arbitrary $\omega$ when $T \to \infty$, combining Eqs. (B5) and (B6) gives the following relation:

$$\lim_{T\to\infty}\int_{-T}^T dt\ F_\text{e}(t)e^{i\omega t} \cong \int_{-\infty}^\infty dt\ F(t)e^{i\omega t}. \tag{B7}$$

Having taken the limit $T \to \infty$, we make the assumption of ergodicity, so the brackets are removed from $F_\text{e}(t)$ on the left-hand side of Eq. (B7). The equation (B7) corresponds to the WKT equation (2); the left-hand side and right-hand side are the power spectral density $I(\omega)$ and the Fourier transform of the ACF $F(\omega)$, respectively.

Based on the discussion above, the WKT requires the convergences of the following Fourier integrals for arbitrary $\omega$:

$$\int_0^\infty dt\ F(\pm t)e^{\pm i\omega t} \quad \text{and} \quad \int_0^\infty dt\ tF(\pm t)e^{\pm i\omega t} . \tag{B8}$$

This requirement is the same as that given in Ref. 9. The second terms omitted in Eqs. (B5) and (B6) come from the highlighted areas outlined by dashed lines in Figs. A1 and A2. Thus, eliminating the unnecessary terms translates into ignoring these highlighted areas.

## APPENDIX C: CONVERGENCE VALUE OF THE SUMMATION IN EQ. (39) FOR THE DEBYE RELAXATION MODEL

As written in Sec. V, knowing the convergence value of the summation in Eq. (39) is difficult without using a specific model for the ACF $F(t_k) = \langle f(t_k)f(0)\rangle$. In this Appendix, we assume as an example that the ACF $F(t_k)$ can be reproduced by the Debye relaxation function, and we confirm that the summation converges to a finite negative value.

The discrete Fourier-Laplace transformation for $F(t_k)$ is written as

$$F^+(\omega_n) = \Delta t \sum_{k=-M/2+1}^{M/2} \theta_k F(t_k) e^{i\omega_n t_k} = \Delta t \left( \sum_{k=0}^{M/2} F(t_k) e^{i\omega_n t_k} - \frac{1}{2} \right), \quad \text{(C1)}$$

where $\theta_k$ is the discretized unit step function of Eq. (32). Differentiating Eq. (C1) with respect to $\omega_n$ gives

$$\frac{dF^+(\omega_n)}{d\omega_n} = i\Delta t \sum_{k=0}^{M/2} F(t_k) t_k e^{i\omega_n t_k} . \quad \text{(C2)}$$

When setting $\omega_n = \pi/\Delta t$ in Eq. (C2), the summation on the right-hand side of Eq. (C2) corresponds to the summation in Eq. (39).

Next, we assume that the ACF can be written by the Debye relaxation function with a relaxation time $\tau$ as $F(t_k) = e^{-t_k/\tau}$, and we substitute this into Eq. (C1):

$$F^+(\omega_n) = \Delta t \left( \sum_{k=0}^{M/2} e^{-t_k/\tau} e^{i\omega_n t_k} - \frac{1}{2} \right) = \Delta t \left( \sum_{k=0}^{M/2} e^{zk} - \frac{1}{2} \right)$$

$$= \Delta t \left( \frac{1 - e^{z(M/2+1)}}{1 - e^z} - \frac{1}{2} \right), \quad \text{(C3)}$$

where $t_k = k\Delta t$, and $z$ is

$$z = (-1/\tau + i\omega_n)\Delta t .\tag{C4}$$

In Eq. (C3), we see $F^+(\omega_n) = 0$ at the Nyquist frequency $\omega_n = \pi/\Delta t$ for $\Delta t \ll \tau \ll T/2$, where $T = M\Delta t$.

Differentiating Eq. (C3) with respect to $\omega_n$ gives

$$\frac{dF^+(\omega_n)}{d\omega_n} = \frac{dz}{d\omega_n}\frac{d}{dz}\left\{\Delta t\left(\frac{1-e^{z(M/2+1)}}{1-e^z} - \frac{1}{2}\right)\right\}$$

$$= i(\Delta t)^2 \frac{e^z}{(1-e^z)^2}\left[-\left\{\frac{M}{2}(1-e^z)+1\right\}e^{zM/2}+1\right] .\tag{C5}$$

Substituting $\omega_n = \pi/\Delta t$ into Eq. (C5) through Eq. (C4), and combining this with Eq. (C2), we get the following form for the summation in Eq. (39):

$$\sum_{k=0}^{M/2} F(t_k)t_k(-1)^k = \frac{1}{i\Delta t}\left.\frac{dF^+(\omega_n)}{d\omega_n}\right|_{\omega_n=\frac{\pi}{\Delta t}}$$

$$= \Delta t \frac{-e^{-\Delta t/\tau}}{(1+e^{-\Delta t/\tau})^2}\left[-\left\{\frac{M}{2}(1+e^{-\Delta t/\tau})+1\right\}(-1)^{M/2}e^{-M\Delta t/2\tau}+1\right] .\tag{C6}$$

On the right-hand side in Eq. (C6), the first term in the square brackets can be ignored when $\tau \ll T/2$, and we thus obtain

$$\sum_{k=0}^{M/2} F(t_k)t_k(-1)^k \cong \Delta t \frac{-e^{-\Delta t/\tau}}{(1+e^{-\Delta t/\tau})^2} \to -0.25\Delta t , \quad \text{for } \Delta t \ll \tau.\tag{C7}$$

From Eqs. (C6) and (C7), we confirm that the summation in Eq. (39) converges to the finite negative value $-0.25\Delta t$ for $\Delta t \ll \tau \ll T/2$ when using the Debye relaxation model.